\documentclass[12pt]{article}
\usepackage{amsmath}
\usepackage{amssymb}
\usepackage{amsthm}
\usepackage{graphicx}
\usepackage[colorlinks=true, allcolors=blue, linktoc=page]{hyperref}
\usepackage{tabularx}
\usepackage{comment}
\usepackage{ dsfont }
\usepackage{cite}

\newcommand{\oo}{{\mathcal O}}
\newcommand{\pa}{\partial}
\newcommand{\wb}{\bar{w}}
\newcommand{\tb}{\bar{t}}
\newcommand{\zb}{\bar{z}}
\newcommand{\ti}[1]{\tilde{#1}}

\newcommand{\ket}[1]{\ensuremath{\left|#1\right>}}

\begin{document}

\numberwithin{equation}{section}

\begin{titlepage}
\hbox to \hsize{\hspace*{0 cm}\hbox{\tt }\hss
   \hbox{\small{\tt }}}

\vspace{1 cm}

\centerline{\bf \Large The large $N$ limit of OPEs in symmetric orbifold CFTs}
\centerline{\bf \Large  with $\mathcal{N}=(4,4)$ supersymmetry}

\vspace{1 cm}

\vspace{1 cm}
 \centerline{\large
Thomas de Beer$^\dagger$\footnote{tdebeer@physics.utoronto.ca}\,,
Benjamin A. Burrington $^{\star}$\footnote{benjamin.a.burrington@hofstra.edu}\,,
Ian T. Jardine$^\dagger$\footnote{itjardine@hotmail.com}\,,
A.W. Peet$^{\dagger\S}$\footnote{awpeet@physics.utoronto.ca}}

\vspace{0.5cm}

\centerline{\it ${}^\star\!\!$ Department of Physics and Astronomy, Hofstra University, Hempstead, NY 11549, USA}
\centerline{\it ${}^\dagger$Department of Physics, University of Toronto, Toronto, ON M5S 1A7, Canada}
\centerline{\it ${}^\S$Department of Mathematics, University of Toronto, Toronto, ON M5S 2E4, Canada}

\abstract{We explore the OPE of certain twist operators in symmetric product ($S_N$) orbifold CFTs, extending our previous work \cite{Burrington:2018upk} to the case of $\mathcal{N}=(4,4)$ supersymmetry. We consider a class of twist operators related to the chiral primaries by spectral flow parallel to the twist. We conjecture that at large $N$, the OPE of two such operators contains only fields in this class, along with excitations by fractional modes of the superconformal currents.  We provide evidence for this by studying the coincidence limits of two 4-point functions to several non-trivial orders.  We show how the fractional excitations of the twist operators in our restricted class fully reproduce the crossing channels appearing in the coincidence limits of the 4-point functions.}
\end{titlepage}
\tableofcontents

\section{Introduction}

The holographic AdS/CFT correspondence \cite{Maldacena:1997re,Witten:1998qj} has proven to be a remarkably productive tool for investigating big questions of quantum gravity in asymptotically Anti de Sitter (AdS) contexts. While string theoretic realizations were vital to its discovery and early development, AdS/CFT has proved to be much more broadly applicable. In \cite{Heemskerk:2009pn}, the question was addressed of which conformal field theories (CFTs) might admit an AdS Einstein gravity dual; two conditions were conjectured to be necessary. First, the theory must admit a large $N$ limit, which in the case of 2D CFTs implies a large central charge $c$.  Second, there must be a large gap in the spectrum of anomalous dimensions, controlled by some coupling constant. This and later developments have demonstrated that only a small subset of all possible CFTs is holographic. Here, we will focus on cases where the AdS bulk is 3D and the CFT is 2D.

For 2D CFTs, a simple way to allow for a large $N$ limit is to orbifold a large number of seed CFTs. The ordinary tensor product of $N$ seeds (${\mathcal{M}}^N$) has a total central charge proportional to $N$, but it also contains $N$ times as many low-lying states in the spectrum as a single seed CFT, making a large $N$ limit ill-defined. Orbifolding by a subgroup of permutations $G_N\subset S_N$ to construct the theory ${\mathcal{M}}^N/G_N$ eliminates many of the states, by projecting out those that do not respect the group action. This strongly suggests that such orbifold CFTs are interesting prototypical candidates for holographic models \cite{Belin:2014fna,Haehl:2014yla,Benjamin:2015hsa, Belin:2015hwa,Benjamin:2015vkc, Belin:2016yll,Belin:2017jli}. Of particular interest to us are the orbifolds by the symmetric group $S_N$, where more powerful computational tools exist using maps to covering surfaces \cite{Lunin:2000yv, Lunin:2001pw, Burrington:2012yn} (for a comprehensive exposition, see \cite{Avery:2010qw}).

In symmetric product $S_N$ orbifold CFTs, the seed CFT is copied $N$ times, and then the spectrum is restricted to the $S_N$ invariant subsector. In addition to this restriction, the action of orbifolding introduces new sectors in the theory where states have non-trivial twisted boundary conditions.  The natural building blocks to study are the states associated with conjugacy classes of  $n$ cycles, because all group elements of $S_N$ may be written as products of cycles. The simplest operators associated with such states are the bare twist $n$ operators, which join together $n$ CFT copies into one. The twisted states dominate at high energies, intuitively due to the rapid growth of the $S_N$ group. More specifically, the growth of high energy states is Hagedorn, which indicates the $S_N$ orbifold CFT to be dual to a stringy kind of gravity \cite{Keller:2011xi}.  Furthermore, it has been shown that certain permutation subgroups $G_N\subset S_N$ also admit well defined large $N$ limits \cite{Belin:2014fna,Belin:2015hwa}, and when the large $N$ limit exists the spectrum also exhibits a Hagedorn spectrum. To fully analyze $S_N$ orbifold CFTs, one must understand the twisted sectors.

To study the twisted sectors, we make use of the covering space techniques of \cite{Lunin:2000yv, Lunin:2001pw, Burrington:2012yn}, of which we now give a brief reminder. When circling a twist sector operator, fundamental fields return to themselves up to the action of $S_N$, which only acts on copy indices.  When several twist sector operators appear in a correlator, twist selection rules must be obeyed: the product of $S_N$ group elements involved must be the identity. Summing over conjugacy class orbits then produces a combinatoric factor. Consider a fundamental field in some simply connected patch. Moving around in this regular patch does not allow encircling twists. In fact, this patch is copied $s$ times, where $s$ is the total number of copy indices involved in the group elements. Crossing the lines excised between a twist operator and infinity advances from one copy to the next, implying a gluing procedure for the copied patches. Gluing for all the patches produces a covering surface with a single copy of the fundamental field.  The conformal map to the cover has branch points where several sheets come together: these are the images of the twist sector operators lifted to the cover. The map is conformal at all points in the regular patch, and so path integrals on the cover and the base are related via the conformal anomaly through the exponentiated Liouville action, the essential calculating tool in \cite{Lunin:2000yv}. Then, appropriate boundary conditions are set at the loci of excised holes in the cover \cite{Lunin:2000yv}, equivalent to picking which twist sector fields are in the correlator, and the appropriate prescription for filling the holes and including operator insertions is found \cite{Lunin:2000yv, Lunin:2001pw, Burrington:2012yn}.

Another significant motivation for us to study $S_N$ orbifold CFTs is the D1-D5 system, which serves as a model to investigate microscopic physics of black holes \cite{Strominger:1996sh}. The D1-D5 intersection is constructed in type IIB string theory compactified on $S^1 \times T^4$, with $N_1$ D1 branes wrapped around $S^1$ and $N_5$ D5 branes wrapped around $S^1 \times T^4$.  The moduli space of this system with two R-R fluxes has been well studied \cite{Seiberg:1999xz,David:1999ec,Giveon:1998ns,Vafa:1995bm,Dijkgraaf:1996xw,Larsen:1999uk,deBoer:1998ip,Dijkgraaf:1998gf}. In one limit, the model becomes a supergravity with an $AdS_3$ factor, along with associated black hole solutions. Another limit admits a description in terms of a superconformal free orbifold CFT with target space $(T^4)^N/S_N$, where $N = N_1N_5$. For the S-dual case of two NS-NS fluxes, string theory on $AdS_3\times {{M}}$ for compact ${{M}}$ is described by a $SL(2,R)$ WZW model \cite{Maldacena:2000hw,Maldacena:2000kv,Maldacena:2001km}. Recent work on the minimal flux case and its connection to symmetric orbifolds may be found in \cite{Eberhardt:2019niq} and references therein.

Understanding details of how the classical black hole spacetime is emergent from the stringy D1-D5 SCFT requires moving away from the orbifold point in moduli space, towards the supergravity point. One approach towards this goal is to deform the SCFT by a marginal operator in conformal perturbation theory. The pertinent deformation operator for reaching towards the supergravity limit lies in the twist 2 sector \cite{David:1999ec}. A particular quantity of physical interest in this program is the set of anomalous dimensions, which describe how the conformal weights of states in the spectrum change under the deformation. An intermediate step in computing the anomalous dimensions is to figure out how operators mix under the deformation. Using covering space techniques to handle twisted correlation functions the mixing can be computed iteratively. However, in practice this is a very tedious task due to the sheer volume of necessary computations. Work on this approach includes \cite{Hampton:2018ygz,Gaberdiel:2015uca,Burrington:2014yia,Carson:2014ena,Carson:2014yxa,Carson:2014xwa, Burrington:2012yq,Avery:2010vk,Avery:2010er,Avery:2010hs, Avery:2009xr, Avery:2009tu,Pakman:2009mi,Pakman:2009zz,Pakman:2009ab,Gava:2002xb,Gomis:2002qi,David:1999ec}.

In a previous work \cite{Burrington:2017jhh} by three of us, we took a more efficient approach, bypassing a good deal of the computational load mentioned above.  We showed that to first order in conformal perturbation theory, the operator product expansion (OPE) {\em on the cover} between a candidate operator and the deformation operator actually encodes both the structure constants and the full mixing operator. However, we only addressed the case where one operator was in the twist sector, while the other was in the untwisted sector.

To extend that insight, in \cite{Burrington:2018upk} the same group of us considered the OPE between two bare twists in a generic bosonic symmetric product $(S_N)$ orbifold CFT. We conjectured that at leading order in $1/N$, the only operators that show up in the OPE of two bare twist operators are bare twist operators and their excitations by fractional Virasoro modes. Our primary motivation was that at leading order in $1/N$, $n$-point correlation functions of bare twists have a universal behavior \cite{Lunin:2000yv}: they only depend on the length of the twists involved ($n_i$), the total number of CFT copies orbifolded ($N$), and the seed CFT central charge ($c$).  Using the conformal bootstrap, $n$-point functions are constructed from 3-point functions; this strongly suggests that there is a universal quality to the three-point functions used to construct the crossing channels.

We gave evidence for our conjecture by expanding a 4-point function in a coincidence limit to several non-leading orders. This gives the conformal weights and structure constants of the exchanged operators in the coincidence limit. Then, the relevant 3-point functions were constructed from fractional Virasoro operators exciting bare twists, and these operators were shown to completely account for the exchange channels in the 4-point function.  We also argued that this idea can be extended because of the factorized form of the 4-point functions. Consider a 4-point function of bare twists excited by fractional Virasoro operators.  This will have one factor identical to the bare twists, which is universal, multiplied by a covering space calculation resulting from the fractional Virasoro modes, which is also universal. So the 4-point functions of twists excited by fractional Virasoro modes also have universal crossing channels. Therefore, in a stronger form of the conjecture, we argued that the bare twists, along with their fractional Virasoro mode excitations, form a closed subalgebra of the operator algebra at large $N$.

The goal of this paper is to extend the conjecture made in \cite{Burrington:2018upk} to ${\mathcal{N}}=(4,4)$ supersymmetric $S_N$ orbifold CFTs, and provide evidence for it in the context of the D1-D5 orbifold CFT. While the extended conjecture we will formulate is tailored to the $\mathcal{N}=(4,4)$ case, we suspect that other theories with extended supersymmetry will have a similar structure. This warrants further exploration, because of structural differences, e.g.~the $\mathcal{N}=(3,3)$ algebra is not a subalgebra of the $\mathcal{N}=(4,4)$. Even though we have restricted the SUSY algebra, we may consider operators that are constructed algebraically as in \cite{Lunin:2001pw}. Because the leading $1/N$ behavior of correlation functions comes from calculations on the spherical cover \cite{Lunin:2000yv}, we expect that our results are independent of the field representation used to realize the $\mathcal{N}=(4,4)$ SUSY.  We choose to use the $c=6$ free field representation of the $\mathcal{N}=(4,4)$ algebra to match with previous work done on the D1-D5 CFT near the orbifold point.

It is important to note that all computations reported here and in our previous work are done to leading order in a $1/N$ expansion.  The large $N$ counting is critical, particularly for interactions between twist fields. In a correlator $\langle O^1_{q_1\cdots q_{r_1}} O^2_{q_{r_1+1}\cdots q_{r_1+r_2}} \cdots\rangle$, each operator carries a normalization proportional to a power of $N$ fixed by the number of copy indices $q_i$ appearing in it. This gives a fixed overall factor of $N$.  We can partition the correlator by how many indices $q_i$ can be chosen from, say $s$.  This leads to a combinatoric factor $N$ choose $s$, which scales as $N^s$ at large $N$. So the combinatorics favor larger choices of $s$, and leading order in $1/N$ corresponds to the largest possible choice of $s$.  However, if $s$ is too large, the correlator decomposes into disconnected diagrams, or may even be zero by a selection rule.  Thus, $s$ is chosen as large as possible, but not too large. The upshot is that at leading order in $1/N$, the covering space is a sphere \cite{Lunin:2000yv}.  For a detailed argument, see section 2.2 of \cite{Burrington:2018upk}.

We organize the paper as follows. In section \ref{section2}, we give a brief review of aspects of the D1-D5 SCFT, emphasizing the importance of fractional modes of the supercurrents.  We consider the fractional modes of the superconformal currents, and show how this leads to a fractional superconformal algebra defined in the presence of a twist operator (some calculations are left to appendix \ref{Appendix_A}).  We further introduce fractional spectral flow parallel to the twist, and show how this helps organize the calculations of later sections. In section \ref{section3}, we present our calculations of two representative 4-point functions in different coincidence limits.  We show, up to fifth order, that the crossing channels are fully reproduced by operators constructed using only fractional modes of the $\mathcal{N}=(4,4)$ superconformal currents acting on the operator exchanged at zeroth order. We display detailed calculations to third order in the crossing channels in the first example to demonstrate the method, and then simply give the fractional excitations that reproduce the crossing channels in later sections (relegating rather lengthy expressions for fourth and fifth order to appendix \ref{Appendix_B}). Finally, in section \ref{discussion}, we summarize our conjecture, and discuss future directions.

\section{D1-D5 CFT and the orbifold point}\label{section2}

\subsection{D1-D5 CFT and \texorpdfstring{$\mathcal{N}=(4,4)$}{} SUSY}
Let us begin with a lightning review of salient features of the D1-D5 SCFT; see \cite{David:2002wn} for more details. At the orbifold point, the D1-D5 system is described by a free $1+1$ dimensional $\mathcal{N}=(4,4)$ superconformal field theory with target space $(T^4)^{N}/S_{N}$, where $N=N_1N_5$ and $N_{p}$ is the number of D$p$-branes for $p=1,5$. The theory has an $SU(2)_L \times SU(2)_R$ R-symmetry, which we associate with indices $\alpha, \dot{\alpha}$ for the $SU(2)_{L,R}$ doublets, and $i,\dot{i}$ for the triplets.  The symmetry currents of the superconformal algebra have the following OPEs
\begin{align}
T(z)T(0)&= \frac{c}{2z^4}+\frac{2T(0)}{z^2}+\frac{\pa T(0)}{z} + \cdots 
\qquad\qquad\qquad\qquad\qquad\qquad
\label{TTOPE}  \\
J^i(z) J^j(0)&=\frac{c}{12z^2}+\frac{i\epsilon^{ijk}J^k(0)}{z} +\cdots  \label{JJOPE}\\
T(z) J^i(0)&= \frac{J^i(0)}{z^2}+\frac{\pa J^i(0)}{z} +\cdots
\end{align}

\noindent
\begin{align}
G^\alpha(z) {\hat{G}}_\beta(0)&=\frac{2c \delta^\alpha_{\ \beta}}{3 z^3}+\frac{4 (\sigma^{i*})^\alpha_{\ \beta} J^i(0)}{z^2}+\frac{2(\sigma^{i*})^\alpha_{\ \beta}\pa J^i(0)}{z}+\frac{2T(0)\delta^a_{\ b}}{z}+\cdots \\
T(z)G^\alpha(0)&=\frac{3G^\alpha(0)}{2z^2}+\frac{\pa G^\alpha(0)}{z} +\cdots \\
T(z)\hat{G}_\beta(0)&= \frac{3\hat{G}_\beta(0)}{2z^2}+\frac{\pa \hat{G}_\beta(0)}{z}+\cdots \\
J^i(z)G^\alpha(0)&= \frac{(\sigma^{i*})^\alpha_{\ \beta} G^\beta(0)}{2z}+\cdots\\
J^i(z)\hat{G}_\alpha(0)&= -\frac{\hat{G}_\beta(0)(\sigma^{i*})^\beta_{\ \alpha}}{2z} +\cdots \,. \label{JGhatOPE}
\end{align}
We consider the `untwisted' case, in the language of Schwimmer and Seiberg \cite{Schwimmer:1986mf}, where the currents are periodic (or antiperiodic for the supercurrents in the Ramond sector).\footnote{Schwimmer and Seiberg \cite{Schwimmer:1986mf} refer to the other algebras as `twisted' because the boundary conditions are such that the currents return to themselves up to an outer automorphism, which do result in twisted boundary conditions on fundamental fields in free field representations \cite{Yu:1987dh}.  This is distinct from the considerations in our current work: we are considering twisted boundary conditions with respect to the orbifold group $S_N$.  While we twist individual copies together, the global currents are sum over the currents of each individual copy, making an orbifold invariant.  This makes the true symmetry currents `untwisted' in both the orbifold and outer automorphism sense in \cite{Schwimmer:1986mf}.}

In addition to the superconformal symmetries, the D1-D5 system has $SU(2)_1 \times SU(2)_2$ symmetry associated with the four torus directions, to which we associate the indices $A,\dot{A}$ respectively. The holomorphic field content of the seed CFT consists of 4 real bosons ($X_a$), 4 real holomorphic fermions ($\psi_b$), and 4 real antiholomorphic fermions ($\ti{\psi}_b$), which we recombine into four complex bosons, four complex holomorphic fermions, and four complex antiholomorphic fermions
\begin{align}
X_{\dot{A} A}&= \frac{1}{\sqrt{2}}\begin{pmatrix} X_3+iX_4 & X_1-iX_2 \\ X_1+iX_2 & -X_3+iX_4 \end{pmatrix} && \nonumber\\
\begin{pmatrix} \psi^{+\dot{1}} \\ \psi^{-\dot{1}}\end{pmatrix}&=\frac{1}{\sqrt{2}}\begin{pmatrix} \psi_1+i\psi_2 \\ \psi_3+i\psi_4 \end{pmatrix}
&
\begin{pmatrix} \ti{\psi}^{\dot{+}\dot{1}} \\ \ti{\psi}^{\dot{-}\dot{1}}\end{pmatrix} &=\frac{1}{\sqrt{2}}\begin{pmatrix} \ti\psi_1+i\ti\psi_2 \\ \ti\psi_3+i\ti\psi_4 \end{pmatrix}
\nonumber\\
\begin{pmatrix} \psi^{+\dot{2}} \\ \psi^{-\dot{2}}\end{pmatrix}&=\frac{1}{\sqrt{2}}\begin{pmatrix} \psi_3-i\psi_4 \\ -\psi_1+i\psi_2 \end{pmatrix}
&
\begin{pmatrix} \ti{\psi}^{\dot{+}\dot{2}} \\ \ti{\psi}^{\dot{-}\dot{2}}\end{pmatrix}&=\frac{1}{\sqrt{2}}\begin{pmatrix} \ti\psi_3-i\ti\psi_4 \\ -\ti\psi_1+i\ti\psi_2 \end{pmatrix} \,.
\end{align}
These fields obey reality constraints
\begin{align}
(X_{\dot{A} {A}})^{\dagger}&\equiv X^{\dot{A}A}=\frac{1}{\sqrt{2}}\begin{pmatrix} X_3-iX_4 & X_1+iX_2 \\ X_1-iX_2 & -X_3-iX_4 \end{pmatrix}=-\epsilon^{\dot{A}\dot{B}}\epsilon^{AB}X_{\dot{A} A}
\nonumber \\
(\psi^{\alpha \dot{A}})^\dagger&\equiv \psi_{\alpha \dot{A}}=-\epsilon_{\alpha \beta} \epsilon_{\dot{A}\dot{B}} \psi^{\beta \dot{B}} \nonumber\\
(\ti{\psi}^{\dot{\alpha} \dot{A}})^\dagger&\equiv \ti{\psi}_{\dot{\alpha} \dot{A}}=-\epsilon_{\dot{\alpha} \dot{\beta}} \epsilon_{\dot{A}\dot{B}} \ti{\psi}^{\dot{\beta}}
\end{align}
where here and elsewhere, we define $-\epsilon^{+-}=1=\epsilon_{+-}, -\epsilon^{\dot{1}\dot{2}}=1=\epsilon_{\dot{1}\dot{2}}$, etc.
The OPEs of these fundamental fields are given by
\begin{align}
\psi^{\alpha \dot{A}}(z_1)\psi^{\beta \dot{B}}(z_2) &= -\frac{1}{z_{12}}\epsilon^{\alpha \beta} \epsilon^{\dot{A}\dot{B}}+\dots \nonumber\\
\tilde{\psi}^{\dot{\alpha}\dot{A}}(\zb_1)\tilde{\psi}^{\dot{\beta} \dot{B}}(\zb_2) &= -\frac{1}{\zb_{12}}\epsilon^{\dot{\alpha} \dot{\beta}} \epsilon^{\dot{A}\dot{B}}+ \dots \nonumber\\
\pa X_{\dot{A}A }(z_1) \pa X_{\dot{B}B}(z_2) &= \frac{1}{(z_{12})^2}\epsilon_{\dot{A}\dot{B}}\epsilon_{AB}+ \dots
\end{align}
where $z_{12}\equiv z_1-z_2$. The fundamental fields combine into the holomorphic copy of superconformal currents via
\begin{align}
T &= \tfrac{1}{2} \epsilon_{\dot{A}\dot{B}}\epsilon_{AB} \partial X^{\dot{A}A} \partial X^{\dot{B}B}
+ \tfrac{1}{2} \epsilon_{\dot{A}\dot{B}} \epsilon_{\alpha \beta} \psi^{\alpha \dot{A}} \partial \psi^{\beta \dot{B}} \label{eq:currents_T}
\\
J^i &= \tfrac{1}{4} \epsilon_{\dot{A}\dot{B}} \epsilon_{\alpha\beta} \psi^{\alpha\dot{A}} (\sigma^{*i})^\beta_{\ \gamma} \psi^{\gamma\dot{B}}
\\
G^{\alpha A} &= \epsilon_{\dot{A}\dot{B}} \psi^{\alpha \dot{A}} \partial X^{\dot{B}A} \label{eq:currents_J}
\end{align}
and similarly for the antiholomorphic copy of the superconformal currents.  We define the conjugate of the superconformal generators as
\begin{equation}
(G^{\alpha A})^\dagger\equiv G_{\alpha A}=-\epsilon_{\alpha \beta}\epsilon_{AB}G^{\beta B} \,.
\end{equation}
The fermionic currents above are related to those in the OPEs of (\ref{TTOPE})-(\ref{JGhatOPE}) via
\begin{equation}
G^{\alpha}=\sqrt{2}G^{\alpha 1} \qquad \hat{G}_\beta=\sqrt{2}G_{\beta 1}=-\sqrt{2}\epsilon_{\beta \gamma}\epsilon_{12}G^{\gamma 2}=-\sqrt{2}\epsilon_{\beta\gamma} G^{\gamma 2}\,.
\end{equation}
To each of the fundamental fields, an additional copy index $p$ is attached (suppressed above).  All OPEs are diagonal in these indices, and the global current operators (\ref{eq:currents_T})-(\ref{eq:currents_J}) are formed by summing over the copy index in these quadratic combinations.

The action of orbifolding introduces new states in twisted sectors, where boundary conditions are non-trivial up to the orbifold group $S_N$. The lowest-weight operator in the twist-$n$ sector is the bare twist, denoted by $\sigma_n \equiv \sigma_{(123\dots n)}$. As one circles around the insertion of a bare twist in the complex plane,
	\begin{equation}
		\phi_{(1)} \rightarrow \phi_{(2)} \rightarrow \dots \rightarrow \phi_{(n)} \rightarrow \phi_{(1)} \,.
	\end{equation}
Here $\phi_{(p)}$ represents a generic field with copy index $p$. The presence of a twist insertion $\sigma_n(0)$ allows the construction of fractional modes of fields
	\begin{equation}
	\begin{split}
		\phi_{-m/n}\sigma_n(0) = \oint \frac{dz}{2\pi i} \sum_{k=1}^n \phi_i(z) e^{-2\pi im(k-1)/n} z^{h-1-m/n}\sigma_n(0) \,.
	\end{split}
	\end{equation}
The twisted boundary conditions make the integral periodic under $z \rightarrow ze^{2\pi i}$: the non-periodicity of the field combination $\sum_{k=1}^n \phi_i(z) e^{-2\pi im(k-1)/n}$ exactly cancels the non periodicity of $z^{h-1-m/n}$, making the integrand single valued. The same construction is applied to the anti-holomorphic sector as well, and applies equally to fundamental and composite fields, such as the symmetry currents above. When there is no twist field, one can set $n=1$ in the above equation to recover the usual mode decomposition for the fields on a single copy.

Fractional modes allow the construction of super (anti)-chiral primary operators in the twist $n$ sector. Chiral primary operators have equal conformal weight and R-charge, $h=q$, whereas anti-chiral primary operators have $h=-q$. The (anti)-chiral primary operators are the holographic dual of bulk supergravity modes . Their construction in the twist $n$ sector follows by applying a series of fractional R-currents on the bare twist. The lowest-weight chiral primaries are \cite{Lunin:2001pw}
	\begin{equation}
		\oo_n =
		\begin{cases}
		&J_{-\frac{n-2}{n}}^+ J_{-\frac{n-4}{n}}^+ \dots J_{-\frac{1}{n}}^+
		\tilde{J}_{-\frac{n-2}{n}}^+ \tilde{J}_{-\frac{n-4}{n}}^+ \dots \tilde{J}_{-\frac{1}{n}}^+ \sigma_n
		 \qquad \ \ n\text{ odd}\\[8pt]
		&J_{-\frac{n-2}{n}}^+ J_{-\frac{n-4}{n}}^+ \dots J_{-\frac{2}{n}}^+
		\tilde{J}_{-\frac{n-2}{n}}^+ \tilde{J}_{-\frac{n-4}{n}}^+ \dots \tilde{J}_{-\frac{2}{n}}^+ \sigma^{+\dot{+}}_n
		 \qquad n\text{ even}\\
	\end{cases}
	\label{eq:chiral_primary}
	\end{equation}
with $h=q=(n-1)/2$. The anti-chiral primaries $\oo^\dagger_n$ are identical but with minus signs replacing all plus signs. Here $\sigma_n^{\alpha\dot{\alpha}}$ denotes the bare twist $n$ operator dressed with a spin field $\mathcal{S}^{\alpha\dot{\alpha}}$ which ensures that the weight $1/2$ fermionic fields do not pick up a negative sign when encircling a twist insertion. We will also use the higher weight chiral primary operators
	\begin{equation}\label{higherwtCP}
	\oo'_n \equiv J^{+}_{-1}\tilde{J}^{+}_{-1} \oo_n
	\end{equation}	
which have $h=q=(n+1)/2$. One could apply the $J^+_{-1}$ and $\tilde{J}^+_{-1}$ independently, but we will not have use for these operators.  Unless specified otherwise, chiral primary will refer to (\ref{eq:chiral_primary}).

The marginal deformation operator of the free orbifold CFT that was mentioned in the introduction takes the form
	\begin{equation}
\label{eq:marginal_deformation_operator}
		\mathcal{O}_D = \epsilon_{\alpha\beta} \epsilon_{\dot{\alpha}\dot{\beta}} \epsilon_{AB} G_{-1/2}^{\alpha A} \tilde{G}_{-1/2}^{\dot{\alpha}B} \sigma_2^{\beta\dot{\beta}}.
	\end{equation}
There are four marginal deformation operators, all in the twist $2$ sector;  (\ref{eq:marginal_deformation_operator}) is a singlet of the various $SU(2)$ symmetries, which makes it easier to handle than the triplet formed by the other three. The fact that this ${\mathcal{O}}_D$ is built out of $G_{-1/2}$ and ${\tilde{G}}_{-1/2}$ acting on $\sigma_2$ makes it twist 2, and this is what motivates us to later focus on OPEs between operators with twist $n$ and twist $2$.

\subsection{Fractional superconformal algebra}

We recall from our previous work \cite{Burrington:2018upk} (independently shown in \cite{Roumpedakis:2018tdb}) that there is a well defined notion of fractional Virasoro operators in the presence of a twist operator.  Near a twist operator associated with an $n$-cycle, the fractional modes of the stress energy tensor obey the following fractional Virasoro algebra
\begin{equation}\label{fractionalalgebra}
\left[L_{\frac{\ell}{n}}, L_{\frac{\ell'}{n}}\right]=\frac{k-k'}{n}L_{\frac{k+k'}{n}} +\delta_{k+k',0}\,\frac{cn}{12}\left(\left(\frac{k}{n}\right)^2-1\right)\frac{k}{n}\,.
\end{equation}
When deriving this result, only component copies of the stress tensor parallel to the twist were used.  This restriction is, of course, not orbifold invariant, and the orbifold invariance is only restored by summing over orbifold images.  Nevertheless, the above algebra gives a technique to construct non gauge-invariant building blocks which are available in any (bosonic) orbifold CFT. One immediately suspects that this should generalize to the superconformal case, where the current algebra is enlarged.

We may use the OPEs of the current algebra (\ref{TTOPE})-(\ref{JGhatOPE}) to find the algebra of fractional modes near a cyclic twist of order $n$. Two example calculations are given in Appendix \ref{Appendix_A}. The full reults are
\begin{align}
\left[L_{\frac{\ell}{n}}, L_{\frac{\ell'}{n}}\right]&=\frac{\ell-\ell'}{n}L_{\frac{\ell+\ell'}{n}}+ \delta_{\ell+\ell',0}\,\frac{cn}{12}\left(\left(\frac{\ell}{n}\right)^2-1\right)\frac{\ell}{n}  \label{LLcommute}\\
\left[J^i_{\ell/n},J^j_{\ell'/n}\right]&=i \epsilon^{ijk} J^k_{\frac{\ell+\ell'}{n}}+\frac{\ell}{n} \frac{ cn}{12} \delta_{\frac{\ell+\ell'}{n},0}\delta^{ij} \\
\left[L_{\frac{\ell}{n}},J^{i}_{\frac{\ell'}{n}}\right]&=-\frac{\ell'}{n} J^i_{\frac{\ell'+\ell}{n}} \\
\left[L_{\frac{\ell}{n}},G^{\alpha}_{\frac{\ell'}{n}}\right]&=\left(\frac{\ell}{2n}-\frac{\ell'}{n}\right) G^{\alpha}_{\frac{\ell+\ell'}{n}}\\
\left[L_{\frac{\ell}{n}},\hat{G}_{\alpha,\frac{\ell'}{n}}\right]&=\left(\frac{\ell}{2n}-\frac{\ell'}{n}\right) \hat{G}_{\alpha,\frac{\ell+\ell'}{n}}\\
\left[ J^{i}_{\frac{\ell}{n}}, G^\alpha_{\frac{\ell'}{n}}\right]&=\frac{1}{2} \left(\sigma^{i*}\right)^\alpha_{\ \beta} G^{\beta}_{\frac{\ell+\ell'}{n}}\\
\left[ J^{i}_{\frac{\ell}{n}}, \hat{G}_{\alpha,\frac{\ell'}{n}}\right]&=-\frac{1}{2} \hat{G}_{\beta,\frac{\ell+\ell'}{n}}\left(\sigma^{i*}\right)^\beta_{\ \alpha} \\
\left\{ G^{\alpha}_{\frac{\ell}{n}}, \hat{G}_{\beta,\frac{\ell'}{n}}\right\}&=2\delta^{\alpha}_{\ \beta} L_{\frac{\ell+\ell'}{n}}+2 \left(\frac{\ell}{n}-\frac{\ell'}{n}\right)(\sigma^{i*})^\alpha_{\ \beta} J^i_{\frac{\ell+\ell'}{n}} \label{GGcommute}
\\ &\qquad  + \frac{nc}{3} \left(\frac{\ell}{n}+\frac{1}{2}\right)\left(\frac{\ell}{n}-\frac{1}{2}\right)\delta^\alpha_{\ \beta} \delta_{\frac{\ell+\ell'}{n},0}\nonumber \,.
\end{align}
We may switch to a $\pm$ basis, introducing the notation
\begin{align}
J^\pm_{\ell/n}&=J^1_{\ell/n}\pm iJ^2_{\ell/n} &
J_{\pm, \ell/n}&=\frac{1}{2}\left(J^1_{\ell/n}\mp iJ^2_{\ell/n} \right) \nonumber\\
\sigma^{\pm*}&=\sigma^{1*}\pm i \sigma^{2*} &
\sigma_{\pm*}&= \frac{1}{2}\left(\sigma^{1*}\mp i \sigma^{2*}\right) \,.
\end{align}
(Note that triplet up indices are now different from down indices, which shows up only in the $G,G$ anticommutators.)

In the $\pm$ basis, the mode algebra is
\begin{align}
\left[L_{\frac{\ell}{n}}, L_{\frac{\ell'}{n}}\right]&=\frac{\ell-\ell'}{n}L_{\frac{\ell+\ell'}{n}}+ \delta_{\ell+\ell',0}\,\frac{cn}{12}\left(\left(\frac{\ell}{n}\right)^2-1\right)\frac{\ell}{n} \label{LLcommutatorPM}\\
\left[J^3_{\ell/n}, J^{\pm}_{\ell'/n}\right]&= \pm J^\pm_{(\ell+\ell')/n} \\
\left[J^+_{\ell/n},J^+_{\ell'/n}\right]&=0 \qquad\qquad\qquad\qquad\qquad\qquad\qquad\ \
	\left[J^-_{\ell/n},J^-_{\ell'/n}\right]=0 \\
\left[J^+_{\ell/n},J^-_{\ell'/n}\right]&=2 J^3_{(\ell+\ell')/n}+2 \frac{\ell}{n} \frac{cn}{12} \delta_{(\ell+\ell')/n,0}\qquad\qquad
\left[J^3_{\ell/n},J^3_{\ell'/n}\right]=\frac{\ell}{n} \frac{cn}{12} \delta_{(\ell+\ell')/n,0}\\
\left[L_{\frac{\ell}{n}},J^{i}_{\frac{\ell'}{n}}\right]&=-\frac{\ell'}{n} J^i_{\frac{\ell'+\ell}{n}}
\end{align}

\noindent
\begin{align}
\left[L_{\frac{\ell}{n}},G^{\alpha}_{\frac{\ell'}{n}}\right]&=\left(\frac{\ell}{2n}-\frac{\ell'}{n}\right) G^{\alpha}_{\frac{\ell+\ell'}{n}}\\
\left[L_{\frac{\ell}{n}},\hat{G}_{\beta,\frac{\ell'}{n}}\right]&=\left(\frac{\ell}{2n}-\frac{\ell'}{n}\right) \hat{G}_{\beta,\frac{\ell+\ell'}{n}}\\
\left[ J^{i}_{\frac{\ell}{n}}, G^\alpha_{\frac{\ell'}{n}}\right]&=\frac{1}{2} \left(\sigma^{i*}\right)^\alpha_{\ \beta} G^{\beta}_{\frac{\ell+\ell'}{n}}\\
\left[ J^{i}_{\frac{\ell}{n}}, \hat{G}_{\alpha,\frac{\ell'}{n}}\right]&=-\frac{1}{2} \hat{G}_{\beta,\frac{\ell+\ell'}{n}}\left(\sigma^{i*}\right)^\beta_{\ \alpha} \\
\left\{ G^{\alpha}_{\frac{\ell}{n}}, \hat{G}_{\beta,\frac{\ell'}{n}}\right\}&=2\delta^{\alpha}_{\ \beta} L_{\frac{\ell+\ell'}{n}}+2 \left(\frac{\ell}{n}-\frac{\ell'}{n}\right)(\sigma_{i*})^\alpha_{\ \beta} J^i_{\frac{\ell+\ell'}{n}}
+ \frac{nc}{3} \left(\frac{\ell}{n}+\frac{1}{2}\right)\left(\frac{\ell}{n}-\frac{1}{2}\right)\delta^\alpha_{\ \beta} \delta_{\frac{\ell+\ell'}{n},0}\,.
\label{GGcommutatorPM}
\end{align}
In the calculations to be presented in section \ref{section3}, we will mainly use the commutation relations involving modes of the stress tensor $T$ and R-current $J$.

\subsection{Fractional spectral flow}

In our previous work \cite{Burrington:2018upk}, the primary fractionally moded Virasoro excitations of bare twists were shown to account for crossing channels of certain four-point functions of bare twists at large $N$.  We conjectured further that the fractional Virasoro mode excitations of the bare twists formed a closed subsector of the operator algebra, i.e., the four point functions of the fractional Virasoro excitations of bare twists have crossing channels that are completely accounted for in this closed subsector.  We argued this on the basis of the universality of the four-point function of bare twists (and therefore their fractional excitations) at large $N$ \cite{Lunin:2000yv}, along with our calculations which were performed to several non-trivial orders. In our current work we wish to extend this conjecture to the supersymmetric case.

In cases with extended superconformal symmetry, the algebra of operators has an additional structure that can be physically useful: spectral flow.  It is natural to ask if there is a notion of spectral flow parallel to the twist that may help us organize our calculations here.  We will see that this is the case. First, we recall that spectral flow of the algebra affects the boundary conditions of the currents
\begin{equation}
J^\pm(z)=e^{\pm 2\pi i \xi} J^\pm(e^{2\pi i}z) \qquad G^+(z)=-e^{i\pi\xi}G^+(e^{2\pi i}z) \qquad G^-(z)=-e^{-i\pi\xi}G^-(e^{2\pi i}z)
\end{equation}
where above we have assumed that the $\xi=0$ case is the Ramond sector.  Note that the spectral flow shift of phase is directly proportional to the R-charge; $G^+$ has charge 1/2 in our conventions.  In terms of the algebra, this corresponds to a shift of
\begin{align}
L'_{\ell}&=L_\ell-\xi J^3_\ell+\frac{c_{\rm tot}}{24}\xi^2\delta_{\ell,0} &
{J}^{\prime 3}_{\ell}&=J^3_{\ell}-\frac{\xi c_{\rm tot}}{12}\delta_{\ell,0} & {G}^{\prime +}_{\ell+\xi/2}&=G^+_{\ell} \nonumber\\
& & {J}^{\prime \pm}_{\ell\pm \xi}&=J^\pm_\ell & {G}^{\prime -}_{\ell-\xi/2}&=G^-_{\ell} \label{flowOps} \,.
\end{align}
The primed operators satisfy the same algebra as the unprimed operators.  Usually one restricts to the case where $\xi=\kappa$ for some integer value $\kappa$ so that the R currents $J^i$ are single valued fields.

We now note that the above shift extends trivially to the fractionally moded algebra of the previous subsection, i.e., setting $c_{\rm tot}=nc$ in (\ref{flowOps}) gives
	\begin{align}
		L'_{\ell/n}&=L_{\ell/n}-\xi J^3_{\ell/n}+\frac{nc}{24}\xi^2\delta_{\ell/n,0} &
		{J'}^3_{\ell/n}&=J^3_{\ell/n}-\frac{\xi n c}{12}\delta_{\ell/n,0} & {G}^{\prime+}_{{\ell/n}+\xi/2}&=G^+_{\ell/n}\nonumber\\
		& & {J^{\prime \pm}_{\ell/n\pm \xi}} &=J^\pm_{\ell/n} & {G^{\prime -}_{\ell/n-\xi/2}} &=G^-_{\ell/n}\,.\label{fractionalshift}
	\end{align}
Again, the primed and unprimed generators satisfy the same algebra (\ref{LLcommutatorPM})-(\ref{GGcommutatorPM}).  In this case, shifts $\xi=1/n$ result in R-currents which are not single valued.  However, near a twist operator there are already currents which are not single valued, such as
\begin{equation}
(J^i)^{\delta/n}=\sum_{p=1}^n (J^i)^{(p)} e^{2\pi i (p) \delta/n}\,.
\end{equation}
If we take the action of spectral flow to be diagonal on the copy indices, then under the spectral flow shift of $\xi=\kappa/n$ for integer $\kappa$, the field $(J^i)^{\kappa/n}$ becomes single valued.  Thus, there seems to be some sense in which a fractional spectral flow can be applied.
The idea of a fractional spectral flow operator was introduced earlier in \cite{Martinec:2001cf}, in the context of $\mathbb{Z}_N$ orbifolds of $AdS_3$. 
See also the more recent \cite{Giusto:2012yz, Chakrabarty:2015foa, Bena:2016agb} regarding the interpretation of fractionally spectrally flowed states in geometric duals. 
Again, in our $S_N$ context, we note that singling out a set of indices $1\cdots n$ is not orbifold invariant.  As with the fractional Virasoro algebra in \cite{Burrington:2018upk}, we use this notion of fractional spectral flow to generate and organize excitations on a certain block, which is then summed over orbifold images to construct an orbifold invariant operator. 

\begin{figure}[ht]
\begin{center}
\includegraphics[width=0.6 \textwidth]{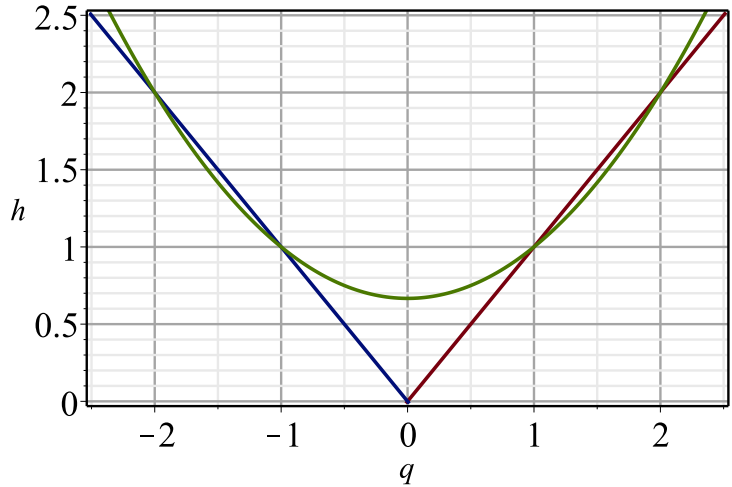}
\end{center}
\vskip-0.3truein
\caption{\small The parabola of conformal weights ($h$) of operators related to the bare twist as a function of charge ($q$) for the case $n=3$.  One can see the bare twist at $q=0$, as well as the chiral primaries at $q=1,2$ which intersect the line $h=q$.  The anti chiral primaries are likewise given by the intersection of the parabola with $h=-q$ at $q=-1,-2$.  Note that in the windows $q\in (1,2)$ and $q=(-2,-1)$ these operators violate the $h\geq |q|$ bound necessary to have unitary representations of superconformal theories.  This window is excluded if one allows spectral flow only in steps $\xi=1/3$, i.e., $\xi=1/n$, from the bare twist. These are exactly the step sizes needed to ensure that the $(J^i)^{\kappa/n}$ fields are single valued. }
\label{fig:parabola}
\end{figure}

To further our discussion, let us ask which operators are related by spectral flow to the bare twist.  To do so, we specialize to the case $c=6$.  We recall that the bare twist has a conformal dimension and charge
\begin{equation}
L_0 \ket{\sigma_n} =h\ket{\sigma_n}=\frac{1}{4}\left(n-\frac{1}{n}\right)\ket{\sigma_n} \qquad J^3_0 \ket{\sigma_n} =q\ket{\sigma_n}=0 \,.
\end{equation}
Under the action of spectral flow parallel to the twist, the new conformal dimension and charge are
\begin{equation}
L'_0 \ket{\sigma_n}_\xi =h'\ket{\sigma_n}_\xi=\left(\frac{1}{4}\left(n-\frac{1}{n}\right)+\frac{n}{4}\xi^2\right)\ket{\sigma_n}_\xi 
\qquad
{J'_0}^3 \ket{\sigma_n}_\xi  =q'\ket{\sigma_n}_\xi=-\frac{\xi n}{2}\ket{\sigma_n}_\xi \,.
\end{equation}
One can easily guess a set of operators that have the above conformal dimension and charge
\begin{equation}
\mathcal{O}_{\xi}\rightarrow A e^{-i\xi\frac{n}{2}\left(\phi_5-\phi_6\right)}
\end{equation}
where $\phi_5$ and $\phi_6$ are the bosonized fermions on the cover
\begin{equation}
\psi^{+\dot{1}}=e^{-i\phi^6} \qquad \psi^{+\dot{2}}=e^{i\phi^5}
\label{eq:bosonization}
\end{equation}
and $A$ is some constant. In (\ref{eq:bosonization}) we have suppressed cocycles, which are important to enforce anticommutation relations, since they do not affect our arguments here. This will be justified in greater detail in section \ref{section3.1}. We see immediately that the chiral primaries $\oo_n,\oo'_n$ are in this family by setting $\xi=-\frac{n-1}{n},-\frac{n+1}{n}$, as are the anti chiral primaries by setting $\xi=\frac{n-1}{n},\frac{n+1}{n}$. See Figure \ref{fig:parabola}.

Interestingly, this spectral flow story gives relations that are not immediately obvious.  For example, we recall that $L_{-1/n}$ lifts to $L^\uparrow_{-1}$ on the cover.  The bare twist lifts to a $\mathds{1}$ insertion, and using the Ward identity to express $L^\uparrow_{-1}$ as a derivative, we see that $\pa_t \mathds{1} =0$.  This gives
\begin{equation}
L_{-1/n}\ket{ \sigma_{n}}=0
\end{equation}
where $\sigma_{n}$ is the bare twist.  However, we may look at this equation after spectral flow to find a statement about chiral primaries,
\begin{equation}
L'_{-1/n}\ket{\sigma_{n}}_{\xi = {\frac{1}{2}}(n\pm1)}=0
\end{equation}
where
\begin{equation}
L'_{-1/n}= L_{-1/n}-\frac{n\pm1}{n} J^3_{-1/n}
\end{equation}
and $\ket{\sigma_{n}}_{\xi = {\frac{1}{2}}(n\pm1)}$ are the two chiral primary fields with $h=q=({n\pm1})/{2}$. These are our friends $\oo'_n$ and $\oo_n$ respectively.  (Recall that the prime on $\oo'_n$ does not refer to spectral flow; instead, it signals the higher weight chiral primary in (\ref{higherwtCP}), c.f. (\ref{eq:chiral_primary}).) The above statement can be verified via direct calculation.  Therefore, many of the excitations one could write are in fact zero when applied to the chiral primary fields, which is easiest to see in the primed basis.  This is actually the case for the full family of operators that lift to $A\exp(-i\xi(\phi_5-\phi_6)/2)$, because all of these are related to the bare twist.  We believe that this simplification in the representation theory results from the fact that the family of operators that lift to $A\exp(-i\xi(\phi_5-\phi_6)/2)$ are the lowest weight operators in their twist and charge class.  The bare twist is simply the $q=0$ case.

Thus, it is convenient to use the shifted basis
	\begin{align}
		L'_{\ell/n}&=L_{\ell/n}-\xi J^3_{\ell/n}+\frac{n}{4}\xi^2\delta_{\ell/n,0} & {J}^{\prime 3}_{\ell/n}&=J^3_{\ell/n}-\frac{\xi n}{2}\delta_{\ell/n,0} & {G}^{\prime +}_{{\ell/n}-\xi/2}&=G^+_{\ell/n}\nonumber	\\
		& & J^{\prime \pm}_{\ell/n\pm\xi}&=J^\pm_{\ell/n} & G^{\prime -}_{\ell/n+\xi/2}&=G^-_{\ell/n}\label{flowbasis}
	\end{align}
which simplifies matters by making it easy to identify operators which evaluate to 0 when acting on the state associated with the lifted operator $A\exp(i\xi n(\phi_5-\phi_6)/2)$. For us, $\xi$ will be determined by the leading order operator exchanged in the OPE. As we investigate various OPEs, the value of $\xi$ will vary.

When multiple fractional modes act to excite a state $\ket{\phi}$, we will use the following convention
	\begin{equation}
		J^{\prime 3}_{-j_1/n} J^{\prime 3}_{-j_2/n} \dots J^{\prime 3}_{-j_a/n} L'_{-\ell_1/n}L'_{-\ell_2/n} \dots L'_{-\ell_b/n} \ket{\phi}
	\end{equation}
such that
	\begin{align}
	& j_1 \geq j_2 \geq \dots \geq j_a \nonumber\\
	& \ell_1 \geq \ell_2 \geq \dots \geq \ell_b \,.
	\end{align}
That is, we always act first with Virasoro modes, in decreasing order of conformal weight, and then we act with R-current modes, also in decreasing order. This convention can be chosen without loss of generality, since any string of modes can be put in such a form using the commutation relations (\ref{LLcommutatorPM})-(\ref{GGcommutatorPM}). Similar choices can be made for the charged R-currents and SUSY modes. The convention above will suffice for the calculations we present in this work.

\section{Crossing channels from three-point functions}\label{section3}

\subsection{Three-point and four-point function generalities}
\label{section3.1}

We start off by thinking about four point functions of operators in the twist sector. Recall from \cite{Burrington:2018upk, Lunin:2000yv, Lunin:2001pw} that different families of exchange channels result from different limits of the same four point function. For our cases of interest, these correspond to the fusions $\sigma_2 \sigma_n\rightarrow \sigma_{n+1}$ and $\sigma_2 \sigma_n\rightarrow \sigma_{n-1}$, which we must treat independently.  We now turn to setting up the general calculation that we will do for each fusion.

In all that follows, we will be focusing on four point functions with the same twist structure as
\begin{equation}
\langle\sigma_n(\infty,\infty)\sigma_2(1,1)\sigma_2(w,\bar{w})\sigma_n(0,0)\rangle\,.
\end{equation}
This twist $n{-}2{-}2{-}n$ structure can be lifted to the covering surface with the map \cite{Lunin:2000yv}
\begin{equation}
z(t)=Ct^n\frac{t-a}{t-1} \label{4ptMap}
\end{equation}
which has the proper branch structure.  The existence of a single complex constant $a$ that parametrizes the map directly corresponds to the single independent cross-ratio for four point functions.  The locations of the ramified points are at $t=0$, $t=\infty$, and $t=t_{\pm}$ with
\begin{equation}
t_\pm=\frac{1}{2n}\left((n-1)a+n+1\pm\sqrt{\left(a-1\right)\left((n-1)^2a-(n+1)^2\right)}\right)\,. \label{tpmEquation}
\end{equation}
The constant $C$ is fixed to
\begin{equation}
C=t_{+}^{-n}\frac{t_+-1}{t_+-a} \label{CEquation}
\end{equation}
such that $z(t_+)=1$ is one of the locations fixed by $SL(2)$ transformations (the others being at $z(0)=0$, and $z(\infty)=\infty$).  The location of the final operator we shall call $w$, and is given by
\begin{equation}
w=z(t_-) \,.
\end{equation}

The Liouville term contribution is the same for all of our four point functions, and is identical to the bare twist four point function.  When normalized by the two point functions, it is given by \cite{Lunin:2000yv}
\begin{align}\label{4ptBare}
&\frac{\langle\sigma_n(\infty,\infty)\sigma_2(1,1)\sigma_2(w,\bar{w})\sigma_n(0,0)\rangle}{\langle\sigma_2(1,1)\sigma_2(0,0)\rangle\langle\sigma_n(\infty,\infty)\sigma_n(0,0)\rangle}\\
&=|C|^{-\frac{c}{4}}|a|^{-\frac{c}{12}\left(\frac{n+1}{2}-\frac{1}{n}\right)}|1-a|^{-\frac{c}{8}}|(n-1)^2a-(n+1)^2|^{-\frac{c}{24}}n^{-\frac{c}{12}}2^{-\frac{5c}{12}}\nonumber
\end{align}
where as usual, the points $z=\infty$ are evaluated by placing at a finite point $z=z_{\infty}$ and then the limit $z_\infty\rightarrow\infty$ is taken, which we implicitly assume henceforth.  We note that in the above expression we have favored writing the function in terms of $a$, which is implicitly a function of $w$ through $w=z(t_-(a))$.  However, we prefer $a$ initially, because in our previous work \cite{Burrington:2018upk}, we find that two fusion possibilities are given by the limits
\begin{align}
& a\rightarrow 0 \qquad \ \, \rightarrow \qquad \sigma_2\sigma_n\rightarrow \sigma_{n+1} \nonumber\\
& a\rightarrow \infty \qquad \rightarrow \qquad \sigma_2\sigma_n\rightarrow \sigma_{n-1} \,. \label{alimits}
\end{align}
We will colloquially refer to these fusion possibilities as twist up and twist down respectively. After evaluating the four point functions in this limit on $a$, we can then turn the series in $a$ into a series in $w$ by inverting the series $w=z(t_-(a))$.

Of course the Liouville term is only part of the calculation.  In general, we will be considering four point functions of the form
\begin{equation}
\frac{\langle \oo_n^\dagger(\infty,\infty) \oo_2^\dagger(1,1) \oo_2(w,\wb) \oo_n(0,0)\rangle}{\langle \oo_n^\dagger (\infty,\infty)\oo_n(0,0)\rangle \langle \oo_2^\dagger(1,1)\oo_2(0,0)\rangle}
\end{equation}
to isolate crossing channels arising from the OPE.  Any four point function with equal twist structure will differ only by the contribution from the covering space calculation, each being multiplied by (\ref{4ptBare}).  In each case, the covering space calculation will be given by
\begin{equation}
\frac{\langle (\oo_n^\dagger)^\uparrow(\infty,\infty) (\oo_2^\dagger)^\uparrow(t_+,\tb_+) \oo^\uparrow_2(t_-,\tb_-) \oo^\uparrow_n(0,0)\rangle}{\langle (\oo^\dagger_n)^\uparrow (\infty,\infty)\oo^\uparrow_n(0,0)\rangle \langle (\oo^\dagger_2)^\uparrow(1,1)\oo^\uparrow_2(0,0)\rangle}\label{4ptLift}
\end{equation}
where the $O^\uparrow$ are the lifted versions of the operators. For us, the lifted operators will always be exponentials, with a proportionality constant that is read off from the map near the branch point. If the map has the form
\begin{equation}
z=z_i+b_i (t-t_i)^{n_i} + \cdots
\end{equation}
then the lifted operator insertion is
\begin{equation}
\oo_n^\uparrow=b_i^{-q^2/n}e^{iq(\phi^5-\phi^6)}\times \text{c.c.} \label{expOperatorForm}
\end{equation}
at finite points.  At the point $t=\infty$, the operator is instead multiplied by $(1/b_i)^{-p^2/n}$, which one can quickly see by mapping to the patch at infinity by $t=1/t'$. One can easily expand the map (\ref{4ptMap}) about the points $t=0,\infty,t_\pm$ to find
\begin{align}
b_0&= Ca \nonumber \\
b_\infty&= C \nonumber\\
b_+&=\frac{Cnt_+^{n-1}(t_+-t_-)}{2(t_+-1)^2} \nonumber \\
b_-&= \frac{Cnt_-^{n-1}(t_--t_+)}{2(t_--1)^2}\,. \label{4ptbs}
\end{align}

The normalizing two point functions can be accounted for with specific maps as well.  In the case that the two point function has insertions at $z=0$ and $z=\infty$, we use the map $z=t^n_i$ giving $b_0=1$ and $b_\infty=1$ for the ramified points at $t=0$ and $t=\infty$.  In the case that the two point function has insertions at $z=0$ and $z=1$, we use $z=\frac{t^n}{t^n-(t-1)^n}$ where $b_0=(-1)^{n+1}$ and $b_1=1$.  In any of these cases, the dressing coefficient of the exponential depends only on $|b|=1$, leaving only the exponentials to contend with, giving pure powers of differences of $t$ positions.  The rest of the normalization, to do with the bare twists, has already been accounted for in (\ref{4ptBare}).

Our aim is to match the crossing channels to the three point functions.  Given the limit in $a$ taken, we know the twist sector, and given the coefficient in the crossing channel mode, we know the conformal weight of the operator.  All that remains is to calculate the three point functions at that conformal weight, and find the correct linear combination of operators.  To do so, we will need the map for the three point functions, which we take to be
\begin{align}
z(t)= \frac{t^{n+1}}{(n+1)t-n} \qquad \mbox{for} \qquad \sigma_2 \sigma_n \rightarrow \label{3ptMapUp} \sigma_{n+1}
\end{align}
and the ramified points are at $0,1,$ and $\infty$.  Again, for the lifted operators we will need the map expanded near the ramified points as
\begin{equation}
z=z_i + a_i(t-t_i)^{n_i}
\end{equation}
and we find
\begin{align}
& a_0=-1/n \nonumber \\
& a_1=\frac{n(n+1)}{2} \nonumber \\
& a_\infty = \frac{1}{n+1}\,.
\end{align}
As with the four-point function, the Liouville term is simply the three point-function of the bare twists, properly normalized, and with the appropriate $t\rightarrow\infty, z\rightarrow \infty$ part removed:
\begin{align}
&\frac{\langle \sigma_n(\infty,\infty)\sigma_2(1,1)\sigma_{n+1}(0,0)\rangle}{\langle \sigma_n(\infty,\infty)\sigma_n(0,0)\rangle}\times
\frac{\langle \sigma_n(1,1)\sigma_n(0,0)\rangle^{1/2}}{\langle \sigma_2(1,1) \sigma_2(0,0)\rangle^{1/2}\langle \sigma_{n+1}(1,1)\sigma_{n+1}(0,0)\rangle^{1/2} } \nonumber \\
&= 2^{-\frac{5}{4}}(n+1)^\frac{2n^2+n+2}{4n}n^{-\frac{2n^2+3n+3}{4(n+1)}} \,. \label{3ptBare}
\end{align}
In practice we used equation (D.81) of \cite{Avery:2010qw} to compute the Liouville contributions.
Again, this bare twist result is universal to all the three point functions we will address, and is multiplied by the covering surface result
\begin{equation}\label{3ptLift}
\frac{\langle \oo^\uparrow_n(\infty,\infty)\oo^\uparrow_2(1,1)\oo^\uparrow_{n+1}(0,0)\rangle}{\langle (\oo^\dagger)^\uparrow_n(\infty,\infty)\oo^\uparrow_n(0,0)\rangle}
\times
\frac{\langle (\oo^\dagger)^\uparrow_n(1,1)\oo^\uparrow_n(0,0)\rangle^{1/2}}{\langle (\oo^\dagger)^\uparrow_2(1,1) \oo^\uparrow_2(0,0)\rangle^{1/2} \langle (\oo^\dagger)^\uparrow_{n+1}(1,1)\oo^\uparrow_{n+1}(0,0)\rangle^{1/2} } \,.
\end{equation}
One may also consider the other possible fusion using the map
\begin{equation}
z(t)= t^{n-1}\left(n-(n-1)t\right) \qquad \mbox{for} \qquad \sigma_2 \sigma_n \rightarrow \sigma_{n-1} \,. \label{3ptMapDown}
\end{equation}
This gives the same result as (\ref{3ptBare}) for the bare twist, replacing $n\rightarrow n-1$, and has a similar covering space calculation (\ref{3ptLift}) replacing the $n+1$ with an $n-1$ at the location $t=0$.  In the map (\ref{3ptMapDown}), the coefficients of the expansion
\begin{align}
z=z_i+a'_i(t-t_i)^{n_i}+\cdots
\end{align}
are given by
	\begin{align}
		a'_0&=n \nonumber\\
		a'_1&= -\frac{n(n-1)}{2}\nonumber\\
		a'_\infty&= -(n-1) \,.\label{eq:twist_down_map_coeffs}
	\end{align}

We make one final digression, and briefly comment on cocycles, and other sources of phases in our calculation.  We recall that in two point functions, the cocycles automatically drop out, since these are of the form $\langle \oo_i^\dagger \oo_i\rangle$, and so the cocycle operators telescope away.  How about for the four point functions of interest? Since these are all of the form
\begin{equation}
\langle \oo_i^\dagger \oo_j^\dagger \oo_j \oo_i\rangle
\end{equation}
they lift to a four point function of the same form
\begin{equation}
\rightarrow \langle (\oo^\dagger_i)^\uparrow (\oo^\dagger_j)^\uparrow \oo^\uparrow_j \oo^\uparrow_i\rangle \,.
\end{equation}
For us, each of our operators lifts to a single exponential, and so only a single cocycle operator dresses each term, and we recall that the daggers reverse the order in which the cocycle operator dresses the main operator.  We take that the operators $\oo^\uparrow_i$ has a covering space momentum $\vec{k}_1$, and is dressed on the left with a cocycle operator.  Hence, we write schematically $\oo^\uparrow_i=e^{c_{\vec{k}_i}} \oo_{\vec{k}_i}$ where $\oo_{\vec{k}_i}$ is the exponential (possibly dressed with excitations).  Further, we may schematically write $(e^{c_{\vec{k}_i}}\oo_{\vec{k}}^\uparrow)^\dagger=\oo_{-\vec{k}}^\uparrow e^{-c_{\vec{k}_i}}$.  Hence, the four point functions of interest have the following form
\begin{align}
&\langle \oo^\uparrow_{-\vec{k}_1} e^{-c_{\vec{k}_1}} \oo^\uparrow_{-\vec{k}_2}e^{-c_{\vec{k}_2}} e^{c_{\vec{k}_2}}\oo^\uparrow_{\vec{k}_2} e^{c_{\vec{k}_1}}\oo^{\uparrow}_{\vec{k}_1}\rangle \nonumber \\
&=\langle \oo^\uparrow_{-\vec{k}_1} \oo^\uparrow_{-\vec{k}_2} \oo^\uparrow_{\vec{k}_2} \oo^{\uparrow}_{\vec{k}_1}\rangle
\end{align}
where the $e^{c_{\vec{k}_2}}$ simply telescope away, and the $e^{c_{\vec{k}_1}}$ give no contribution, since the total momentum of the operator appearing between $e^{c_{\vec{k}_1}}$ and $e^{-c_{\vec{k}_1}}$ is 0.  Hence, the cocycles give no contributions to the four point functions under consideration in our current work, since they are all single exponentials.  In cases where there is more than one exponential, the cocycles will be important for keeping track of relative phases, for example when looking at the deformation operator.

How about the three-point functions of interest?  Using similar arguments to those used above, the general form of the lifted three point functions is
\begin{align}
&\rightarrow \langle \left[\oo^\uparrow_{-\vec{k}_1} e^{-c_{\vec{k}_1}}\right] \left[\oo^\uparrow_{-\vec{k}_2}e^{-c_{\vec{k}_2}}\right] \left[e^{c_{\vec{k}_1+\vec{k}_2}}\oo^\uparrow_{\vec{k}_1+\vec{k}_2}\right]\rangle \nonumber\\
&=\langle \left[\oo^\uparrow_{-\vec{k}_1}\right] \left[e^{-c_{\vec{k}_1}} \oo^\uparrow_{-\vec{k}_2}e^{c_{\vec{k}_1}}\right] \left[\oo^\uparrow_{\vec{k}_1+\vec{k}_2}\right]\rangle \,.
\end{align}
In this case, there is a remaining phase coming from commuting the cocycle operators through $\oo^\uparrow_{\vec{k}_2}$.  However, since this is just an overall phase, we may write
\begin{equation}
\langle \left[\oo^\uparrow_{-\vec{k}_1}\right] \left[e^{-c_{\vec{k}_1}} \oo^\uparrow_{-\vec{k}_2}e^{c_{\vec{k}_1}}\right] \left[\oo^\uparrow_{\vec{k}_1+\vec{k}_2}\right]\rangle=e^{i\theta_{\vec{k}_1,\vec{k}_2}} \langle \left[\oo^\uparrow_{-\vec{k}_1}\right] \left[\oo^\uparrow_{-\vec{k}_2}\right] \left[\oo^\uparrow_{\vec{k}_1+\vec{k}_2}\right]\rangle.
\end{equation}
Thus, properly speaking, we would have to account for this phase using the cocycles.  However, for our current purposes we will be able to ignore such phases, since they cancel in the four point function crossing channels.  To see this, we recall that to reconstruct the four-point function from three-point functions, two OPEs are used: one on the pair of operators, and one on the pair of conjugate operators.  Thus, we have the phase and its complex conjugate that appears in the crossing channel. Schematically, this appears as
\begin{align}
\rightarrow & \langle \underbrace{(\oo^\dagger_i)^\uparrow (\oo^\dagger_j)^\uparrow} \underbrace{ \oo^\uparrow_j \oo^\uparrow_i}\rangle \,. \nonumber\\
& \kern 2em C_{ijk}^* \kern 2em C_{ijk}
\end{align}
In addition to the phase coming from the cocycle, we may also drop overall phases, knowing that they would cancel out when examining the crossing channel in exactly the same way.

We now turn to the specific cases we study in our work. In each of the following subsections we first find the appropriate four point function by evaluating (\ref{4ptLift}), substituting in the appropriate exponentials (\ref{expOperatorForm}), and multiplying by (\ref{4ptBare}).  We then expand this four point function by expanding appropriately in $a$ (\ref{alimits}), and then use $z(t_-)=w$ to convert this to an expansion in $w$.  Knowing the limit in $a$, we know the twist sector of the exchanged field, and knowing the power of $w$ in the expansion we know the weight of the operator.  We construct a space of orthogonal candidate operators using only the fractional algebra (\ref{LLcommutatorPM})-(\ref{GGcommutatorPM}) in the spectrally flowed basis (\ref{flowbasis}), acting on an operator of appropriate twist and charge that is a pure exponential (\ref{expOperatorForm}).  We calculate the three point functions with these candidate operators using (\ref{3ptBare}) and (\ref{3ptLift}), or their analogues for $\sigma_2 \sigma_n \rightarrow \sigma_{n-1}$. The three point functions give the appropriate structure constants, which yield the contributions of candidate operators to the crossing channel.

We find that the operators constructed in this way completely account for the exchange channels in our four-point functions.  Our conjecture is that all operators related to the bare twist by spectral flow, along with their excitations using the fractional modes (\ref{LLcommutatorPM})-(\ref{GGcommutatorPM}), form a closed subsector of the operator algebra at large $N$.

\subsection{Fusing twist \texorpdfstring{$n$}{} and twist \texorpdfstring{$2$}{} chiral primaries (CPs) to twist \texorpdfstring{$n+1$}{}}

We start by considering the fusion of the twist $n$ chiral primary which lifts to
\begin{equation}
\oo_n\rightarrow \oo^\uparrow_n = b^{-\frac{(n-1)^2}{4n}}e^{i\frac{(n-1)}{2}\left(\phi_5-\phi_6\right)}\times \text{c.c.} \label{CPCPTwistUpTwistn}
\end{equation}
with $h=q=\frac{n-1}{2}$, and the twist 2 chiral primary which lifts to
\begin{equation}
\oo^\uparrow_2 = b^{-\frac{1}{8}}e^{\frac{i}{2}\left(\phi_5-\phi_6\right)}\times \text{c.c.} \label{CPCPTwistUpTwist2}
\end{equation}
with $h=q=\frac{1}{2}$.  Since these are both chiral primaries, we expect the OPE to be non-singular, and close on a third chiral primary at leading order, obeying the expected chiral ring structure \cite{Lerche:1989uy} for extended supersymmetry.  We substitute (\ref{CPCPTwistUpTwistn}) and (\ref{CPCPTwistUpTwist2}) into (\ref{4ptLift}) and find
\begin{align}
& \frac{\langle (\oo_n^\dagger)^\uparrow(\infty,\infty) (\oo_2^\dagger)^\uparrow(t_+,\tb_+) \oo^\uparrow_2(t_-,\tb_-) \oo^\uparrow_n(0,0)\rangle}{\langle (\oo^\dagger_n)^\uparrow (\infty,\infty)\oo^\uparrow_n(0,0)\rangle \langle (\oo^\dagger_2)^\uparrow(1,1)\oo^\uparrow_2(0,0)\rangle}
=\lim_{t_\infty\rightarrow \infty} \left|\frac{b_0}{b_\infty}\right|^{-\frac{(n-1)^2}{2n}}\left|b_+ b_-\right|^{-\frac{1}{4}} \nonumber
\\& \quad \times \left|t_\infty\right|^{(n-1)^2}\left|t_\infty-t_+\right|^{(n-1)}
\left|t_\infty-t_-\right|^{-(n-1)}\left|t_\infty\right|^{-(n-1)^2}\left|t_+-t_-\right|^{-1} \nonumber \\
&= \left|\frac{b_0}{b_\infty}\right|^{-\frac{(n-1)^2}{2n}}\left|b_+ b_-\right|^{-\frac{1}{4}}\left|t_+-t_-\right|^{-1}.
\end{align}
The full four point function is then given by the above answer multiplied by (\ref{4ptBare}) at $c=6$, giving
	\begin{equation}
	\begin{split}
		& \frac{\langle \oo_n^\dagger(\infty,\infty) \oo_2^\dagger(1,1) \oo_2(w,\wb) \oo_n(0,0)\rangle}{\langle \oo_n^\dagger (\infty,\infty)\oo_n(0,0)\rangle \langle \oo_2^\dagger(1,1)\oo_2(0,0)\rangle}
= \left|\frac{b_0}{b_\infty}\right|^{-\frac{(n-1)^2}{2n}}\left|b_+ b_-\right|^{-\frac{1}{4}}\left|t_+-t_-\right|^{-1}
|C|^{-\frac{3}{2}} \\
& \qquad \times|a|^{-\frac{1}{2}\left(\frac{n+1}{2}-\frac{1}{n}\right)} |1-a|^{-\frac{3}{4}}|(n-1)^2a-(n+1)^2|^{-\frac{1}{4}}n^{-\frac{1}{2}}2^{-\frac{5}{2}}\,.
\label{eq:CPCP_unexpanded_4pf}
	\end{split}
	\end{equation}
Using the expressions for the $b_i$ in (\ref{4ptbs}), along with (\ref{CEquation}) and (\ref{tpmEquation}), one has the expression written in terms of $a$.  It is now a simple matter to expand near $a\rightarrow 0$, which is the appropriate limit to produce a twist $n+1$ field, and then invert the expansion $w=z(t_-(a))$ to find the following expansion in $w$
\begin{align}
& \frac{\langle \oo_n^\dagger(\infty,\infty) \oo_2^\dagger(1,1) \oo_2(w,\wb) \oo_n(0,0)\rangle}{\langle \oo_n^\dagger (\infty,\infty)\oo_n(0,0)\rangle \langle \oo_2^\dagger(1,1)\oo_2(0,0)\rangle} \nonumber \\
&= \frac{(n+1)}{2n}\bigg[\bigg(1+n^{-\frac{2(n-1)}{n+1}}w^{2/(n+1)}+2(n-1)^2n^{-\frac{2(2n-1)}{(n+1)}}w^{3/(n+1)} \nonumber \\
& \qquad +n^{-\frac{2(3n-1)}{n+1}}\left(3n^4-14n^3+23n^2-14n+3\right)w^{4/(n+1)} \label{CPCPTwistUpExpand}\nonumber\\
&\qquad +n^{-\frac{2(4n-1)}{n+1}}\frac{2(n-1)^2(6n^4-35n^3+67n^2-35n+6)}{3}w^{5/(n+1)}+\cdots \bigg)\times \text{c.c.}\bigg]\,.
\end{align}
This completes the 4-point function calculation for this case.

In the remainder of this section we will follow our conjecture to reproduce the leading order and several non-leading order coefficients of the coincidence limit (\ref{CPCPTwistUpExpand}) through 3-point function calculations. The crossing channels of order $w^{1/(n+1)},w^{2/(n+1)}$ and $w^{3/(n+1)}$ will be discussed in quite some level of detail. To avoid boring the reader, later sections with similar calculations will be presented in less detail.

The leading order coefficient is already explained by the three point function worked out in
\cite{Lunin:2001pw}, where they found that the structure constant of (\ref{CPCPTwistUpTwistn}) with (\ref{CPCPTwistUpTwist2}) and the twist $n+1$ sector field
\begin{equation}
\oo_{n+1}\rightarrow \oo^\uparrow_{n+1} = b^{-\frac{n^2}{4(n+1)}}e^{i\frac{n}{2}\left(\phi_5-\phi_6\right)}\times \text{c.c.} \label{CPCPTwistUpTwistnp1}
\end{equation}
is given by $C_{n,2,n+1}=\frac{n+1}{2n}$, and so squaring this gives the leading order term in (\ref{CPCPTwistUpExpand}).  We expect excitations of $\oo_{n+1}$ to reproduce non-leading crossing channels, to which we now turn.

We begin by noting that the above expansion (\ref{CPCPTwistUpExpand}) begins at order $w^{2/(n+1)}$, and so this excludes the crossing channel from a weight $1/(n+1)$ excitation of the field (\ref{CPCPTwistUpTwistnp1}). We now check this with a quick calculation. Given the leading order exchanged operator (\ref{CPCPTwistUpTwistnp1}), we find $\xi = \frac{n}{n+1}$, and thus the basis (\ref{flowbasis}) becomes
\begin{align}
&L'_{-1/(n+1)}=L_{-1/(n+1)}-\frac{n}{n+1}J^3_{-1/(n+1)} \nonumber \\
&J^{\prime 3}_{\:-1/(n+1)}=J^3_{-1/(n+1)} \,.
\end{align}
We see that there are two natural candidate excitations that remain in the same charge sector
\begin{align}
& L'_{-1/(n+1)} \ket{ \oo_{n+1}}=0 \nonumber \\
& J^{\prime 3}_{\:-1/(n+1)} \ket{ \oo_{n+1}}\neq 0\,.
\end{align}
While the first excitation annihilates the state, the second excitation gives a non-zero operator.  We see that, given the map (\ref{3ptMapUp}), we may construct the lifted operator at the origin (dropping the antiholomorphic side to condense notation)
\begin{align}
{J'}^3_{-1/{(n+1)}} \ket{ \oo_{n+1}} &\rightarrow \oint \frac{dt}{2\pi i} z(t)^{-1/(n+1)} \frac{i}{2}:(\pa\phi_5-\pa \phi_6)(t)::(a_0)^{-\frac{n^2}{4(n+1)}} e^{i\frac{n}{2}\left(\phi_5-\phi_6\right)}: \nonumber \\
& = (a_0)^{-\frac{n^2}{4(n+1)}}\frac{n^{\frac{-n}{n+1}}}{2}:\left(in(\pa\phi_5-\pa\phi_6)-n\right)
e^{i\frac{n}{2}\left(\phi_5-\phi_6\right)}: \,.
\end{align}
Therefore, the relevant lifted three point function is proportional to
\begin{equation}
t_\infty^{\frac{n^2}{2}}\langle \left[e^{-i\frac{n-1}{2}(\phi_5-\phi_6)}(t_\infty)\right] \left[e^{-\frac{i}{2}(\phi_5-\phi_6)}(1)\right]\left[\left(\frac{in}{2}(\pa\phi_5-\pa\phi_6)-\frac{n}{2}\right)
e^{i\frac{n}{2}\left(\phi_5-\phi_6\right)}(0)\right]\rangle=0
\end{equation}
in the limit that $t_\infty\rightarrow \infty$.  Hence, although there is an operator at the relevant level using the fractional algebra, this candidate operator does not show up in the OPE, and so does not contribute to the crossing channel.  We could have inferred this from the absence of the term $w^{1/(n+1)}$ in the expansion directly, however it is good to see this explicitly here.

At the first nontrivial order, we find a space of orthogonal states/operators and their norms
\begin{align}
&L'_{-2/(n+1)}\ket{ \oo_{n+1}} \qquad \qquad |L'_{-2/(n+1)}\ket{ \oo_{n+1}}|^2=\frac{3}{(n+1)^2} \nonumber \\
&J^{\prime 3}_{\: -2/(n+1)}\ket{  \oo_{n+1}} \qquad \qquad |J^{\prime 3}_{\: -2/(n+1)}\ket{ \oo_{n+1}}|^2=1 \nonumber \\
&\left(\left(J^{\prime 3}_{-1/(n+1)}\right)^2-\frac{n+1}{6}L'_{-2/(n+1)}\right)\ket{\oo_{n+1}} \nonumber \\
& \qquad \qquad \qquad \qquad \left|\left(\left(J^{\prime 3}_{-1/(n+1)}\right)^2-\frac{n+1}{6}L'_{-2/(n+1)}\right)\ket{  \oo_{n+1}}\right|^2=\frac{5}{12}
\end{align}
with the remaining excitations involving $L'_{-1/(n+1)}$, which may always be commuted through to the right, giving 0 plus one of the operators listed above\footnote{One could imagine pairs of charged operators appearing as well, such as a $J^+ J^-$ pair, that leave the total charge unchanged.  However, the weight restriction confines us to look at $J^+_{-1/(n+1)}$ and $J^-_{-1/(n+1)}$.  In such a case, we may always write these charged operators furthest to the right.  We then recall that $J^+_{-1/(n+1)}$ increases the charge by 1, but only increases the conformal weight by $1/(n{+}1)$, putting us below the parabola of lowest weight in a given charge sector.  Thus, the only combination that appears must be $J^+_{-1/(n+1)}J^-_{-1/(n+1)}$, which may be replaced with its commutator, yielding an uncharged operator, removing the need to consider such operators.  This argument generalizes for higher weight excitations, until these weights exceed 1, giving an operator of allowed charge, weight, and twist, when acting on the chiral primaries.  The authors of \cite{Lunin:2001pw} described this same phenomenon when constructing the chiral primaries using the fermionic representation of the currents, and pointing out that these fermions fill a Fermi sea, restricting which modes of $J^+$ can operate.}.  Interestingly, we will find that only the first two states contribute.  We find that for the map (\ref{3ptMapUp})
\begin{align}
&L'_{-2/(n+1)} \oo_{n+1}  \rightarrow a_0^{\frac{-n^2}{4(n+1)}}\frac{n^{\frac{1-n}{n+1}}}{4(n+1)}\bigg(4nT -2i n(\pa \phi_5-\pa\phi_6)+n-6\bigg)e^{\frac{in}{2}(\phi_5-\phi_6)} \nonumber \\
&J^{\prime 3}_{\: -2/(n+1)} \oo_{n+1} \rightarrow a_0^{\frac{-n^2}{4(n+1)}}\frac{n^{\frac{1-n}{n+1}}}{2}\bigg(in(\pa^2\phi_5-\pa^2\phi_6)-2i(\pa \phi_5-\pa\phi_6)-n+1\bigg)e^{\frac{in}{2}(\phi_5-\phi_6)} \label{CPCPTwistUpBasis2}
\end{align}
where $T$ is the full stress tensor for the $c=6$ representation of $\mathcal{N}=(4,4)$ superconformal symmetry.

To evaluate the three-point functions, we recall that there exists a quick way to evaluate correlators of dressed exponentials (see for example chapter 6 of \cite{Polchinski:1998rq}).  One replaces each term dressing an exponential by a dressing minus the contractions with other exponentials.  One then takes the correlator of the modified dressings stripped of the exponentials, and multiplies this by the correlator of just the exponentials.  However, the correlator of just the exponentials is common to all terms in the three point functions we are computing, and so is part of the factored term in front of (\ref{CPCPTwistUpExpand}).  To compute just the correction term, then, we need to calculate the expectation value of just the modified dressings to the exponentials, along with the modification to the normalization, given by the norm of the states above.  Furthermore, we are exploring a relatively simple case: only one of the terms in the three point functions we are calculating comes with a dressing, i.e., the general form is
\begin{equation}
\langle \left[e^{-i\frac{(n-1)}{2}(\phi_5-\phi_6)}(t_\infty)\right] \left[e^{-\frac{i}{2}(\phi_5-\phi_6)}(1)\right]\left[({\rm dressing})e^{\frac{in}{2}(\phi_5-\phi_6)}\right]\rangle \,.
\end{equation}
Therefore, we only need to compute the 1 point function of the dressing term, modified by contractions with the other exponentials.  Since this is a one point function, only the constant terms survive: the constant parts of the dressing, along with the contractions with other exponentials.  Furthermore, the contractions of the dressing with the exponential at $t_\infty$ get suppressed in the final limit as $t_\infty\rightarrow \infty$.  Thus, to calculate the corrections, we simply take the dressing, and replace it with the contractions with the exponential at $t=1$.  This gives
\begin{align}
&L'_{-2/(n+1)}\ket{ \oo_{n+1} } \rightarrow_{\rm dc} \frac{-3n^{\frac{1-n}{(n+1)}}}{2(n+1)} \nonumber \\
&J^{\prime 3}_{\:-2/(n+1)}\ket{ \oo_{n+1} } \rightarrow_{\rm dc}  \frac{-n^{\frac{1-n}{n+1}}}{2}
\end{align}
where we use $\rightarrow_{\rm dc}$ to indicate the dressing correction.  We combine the dressing correction and the correction to the normalization $1/\sqrt{{\rm norm}}$ from (\ref{CPCPTwistUpBasis2}), and find
\begin{align}
&L'_{-2/(n+1)}\ket{ \oo_{n+1} } \rightarrow_{\rm c} \frac{-\sqrt{3}n^{\frac{1-n}{(n+1)}}}{2} \nonumber \\
&J^{\prime 3}_{\: -2/(n+1)}\ket{ \oo_{n+1} } \rightarrow_{\rm c}  \frac{-n^{\frac{1-n}{n+1}}}{2}
\end{align}
where by $\rightarrow_{\rm c}$ we mean the total correction.  In the case $n\geq2$, the above operators are Virasoro primary, and so they contribute to the crossing channels by squaring, hence, we add the squares of the above corrections, finding
\begin{equation}
\left(\frac{-\sqrt{3}n^{\frac{1-n}{(n+1)}}}{2}\right)^2+\left(\frac{-n^{\frac{1-n}{n+1}}}{2}\right)^2=n^{\frac{-2(n-1)}{n+1}}
\end{equation}
agreeing with the first correction term in (\ref{CPCPTwistUpExpand}).  Another way of saying this is that the operator associated with the state
\begin{equation}
\left(\frac{-\sqrt{3}n^{\frac{1-n}{(n+1)}}}{2}L'_{-2/(n+1)} +\frac{-n^{\frac{1-n}{n+1}}}{2}{J}^{\prime 3}_{-2/(n+1)}\right)\ket{ \oo_{n+1} } \label{CPCPTwistUpw2Operator}
\end{equation}
is the one that shows up in the OPE of the operators (\ref{CPCPTwistUpTwistn}) and (\ref{CPCPTwistUpTwist2}).

Next up we will do the order $w^{3/(n+1)}$ computation. At this order, the orthogonal set of five non-null operators with their respective norms is:
	\begin{align}
& L'_{-\frac{3}{n+1}} \ket{\oo_{n+1}}\qquad\qquad\qquad\qquad\ \,\left|L'_{-\frac{3}{n+1}}\ket{\oo_{n+1}} \right|^2= \frac{12}{(n+1)^2}  \nonumber\\
& J^{\prime 3}_{-\frac{3}{n+1}} \ket{\oo_{n+1}} \qquad\qquad\qquad\qquad \ \ \left|J'^3_{-\frac{3}{n+1}}\ket{\oo_{n+1}} \right|^2=\frac{3}{2} \nonumber\\
& J'^3_{-\frac{1}{n+1}}L'_{-\frac{2}{n+1}}\ket{\oo_{n+1}} \qquad\qquad\qquad\left|J'^3_{-\frac{1}{n+1}}L'_{-\frac{2}{n+1}}\ket{\oo_{n+1}} \right|^2=\frac{2}{(n+1)^2}  \nonumber\\
& \left(J'^3_{-\frac{2}{n+1}}J'^3_{-\frac{1}{n+1}}+\frac{n+1}{12}L'_{-\frac{3}{n+1}}\right)\ket{\oo_{n+1}}\nonumber
\\ &\qquad\qquad    \left|\left(J'^3_{-\frac{2}{n+1}}J'^3_{-\frac{1}{n+1}}+\frac{n+1}{12}L'_{-\frac{3}{n+1}}\right)\ket{\oo_{n+1}} \right|^2=\frac{5}{12}\nonumber \\
& \left(\left(J'^3_{-\frac{1}{n+1}}\right)^3+\frac{3(n+1)}{8} J'^3_{-\frac{1}{n+1}}L'_{-\frac{2}{n+1}}\right)\ket{\oo_{n+1}} \nonumber
\\ & \qquad \qquad \left|\left(\left(J'^3_{-\frac{1}{n+1}}\right)^3+\frac{3(n+1)}{8} J'^3_{-\frac{1}{n+1}}L'_{-\frac{2}{n+1}}\right)\ket{\oo_{n+1}} \right|^2=\frac{15}{32}\,. \label{eq:operators_3_firstlast}
	\end{align}
Again, only the first two operators are found to contribute, and these lift to
	\begin{align}
L'_{-\frac{3}{n+1}}\oo_{n+1} & \rightarrow a_0^{\frac{n^2}{4(n+1)}} \frac{ n^{\frac{n+4}{n+1}}}{2 n^3 (n+1)}\bigg( 8 n T -2 n^2 T'+ i n^2 (\partial^2 \phi_5-\partial^2 \phi_6)
		 \nonumber\\
& \qquad  + in(n-3) (\partial \phi_5-\partial \phi_6)-n^2+9n-8 \bigg)e^{\frac{in}{2}(\phi_5-\phi_6)}
		\nonumber\\
	 J^{\prime 3}_{-\frac{3}{n+1}} \oo_{n+1} & \rightarrow a_0^{\frac{n^2}{4(n+1)}}\frac{n^{-\frac{3}{n+1}}}{4 n^2} \bigg( i n^2 \left(\partial^3 \phi_5-\partial^3 \phi_6\right)-6 i n \left(\partial^2 \phi_5- \partial^2 \phi_6\right)
	 \nonumber\\
& \qquad- i(3n-6) \left(\partial \phi_5-\partial \phi_6\right) -2 n^2+5 n-2 \bigg)e^{\frac{in}{2}(\phi_5-\phi_6)} \,.
	\end{align}
Putting these in 3-point functions and handling the dressing as explained above, and together with the norms in (\ref{eq:operators_3_firstlast}) we find
\begin{align}
L'_{-\frac{3}{n+1}} \ket{\oo_{n+1}} &\rightarrow_{\text{c}} \frac{2 (n-1) n^{-\frac{n+4}{n+1}}}{\sqrt{3} n^3}\nonumber\\
J^{\prime 3}_{-\frac{3}{n+1}}  \ket{\oo_{n+1}} &\rightarrow_{\text{c}} -\frac{\sqrt{\frac{2}{3}} (n-1) n^{\frac{3}{n+1}}}{n^2} \,.
\end{align}
Adding up the squares indeed gives the order  $w^{3/(n+1)}$ coefficient in (\ref{CPCPTwistUpExpand}). That is, the operator accounting for exchange is
	\begin{equation}
	\begin{split}
		\frac{(n-1) n^{-\frac{2 n - 1}{n+1}}}{\sqrt{3}} \left(\sqrt{2} J^{\prime 3}_{-\frac{3}{n+1}}+2 L'_{-\frac{3}{n+1}} \right)\ket{\oo_{n+1}} \,.
	\end{split}
	\end{equation}
Notice that so far none of the operators involving $J^{\prime 3}_{-1/(n+1)}$ contributed to the exchange channels. It would be interesting to understand exactly why this happens, even though this is an excitation-basis dependent statement.

Following our conjecture, we have also matched the order $w^{4/(n+1)}$ and $w^{5/(n+1)}$ coefficients in (\ref{CPCPTwistUpExpand}), the results of which are given in Appendix \ref{Appendix_B}.

\subsection{Fusing twist \texorpdfstring{$n$}{} and twist \texorpdfstring{$2$}{} CPs to
twist \texorpdfstring{$n-1$}{}}

In this section we consider the other fusion possibility $\sigma_n \sigma_2 \rightarrow \sigma_{n-1}$ of the same four point function. We expand (\ref{eq:CPCP_unexpanded_4pf}) around $a \rightarrow \infty$ and substitute $a(w)$ obtained by series inversion of the defining relation $w(a) = z(t_-)$.
	\begin{align}
		&\frac{\langle \oo^{\dagger}_{n}(\infty,\infty)\oo^{\dagger}_{2}(1,1) \oo_{2}(w,\bar{w})  \oo_{n}(0,0) \rangle}{\langle \oo^{\dagger}_{n}(\infty,\infty) \oo_{n}	(0,0)\rangle\langle \oo^{\dagger}_{2}(1,1) \oo_{2}(0,0)\rangle}\label{eq:twist_down_4pf} \nonumber\\
		&=\left(2n(n-1)\right)^{-2} \bigg[ \bigg(1+2n^{-\frac{2}{n-1}}w^{\frac{1}{n-1}}+n^{-\frac{2+2n}{n-1}}(3n^2+4n+3)w^{\frac{2}{n-1}}\nonumber\\
		&\quad+2n^{-\frac{2+4n}{n-1}}(2n^4+7n^3+11n^2+7n+2)w^{\frac{3}{n-1}}\nonumber\\
		&\quad + \frac{1}{3} n^{-\frac{6 n+2}{n-1}} \left(15 n^6+94 n^5+256 n^4+348 n^3+256 n^2+94 n+15\right)w^{\frac{4}{n-1}}\nonumber\\
		&\quad + \frac{1}{3} n^{-\frac{8 n+2}{n-1}} \left(18 n^8+171 n^7+706 n^6+1561 n^5+2022 n^4+1561 n^3+706 n^2+171 n+18\right)w^{\frac{5}{n-1}}\nonumber\\
		& \quad + \oo(w^{\frac{6}{n-1}})\bigg)\times\text{c.c.} \bigg] \,.
	\end{align}
From this we see that the leading order exchanged operator has $h=q=n/2$, and twist $n-1$. One can easily guess the operator to be
	\begin{equation}
		\oo'_{n-1} = J^{+}_{-1}\tilde{J}^{+}_{-1} \oo_{n-1}
	\end{equation}
which is the non-minimal conformal weight chiral primary. The structure constant associated with this operator correctly matches the leading order coefficient in the expansion above, as will be demonstrated now. Using the map (\ref{3ptMapDown}) the proposed operator lifts to
		\begin{equation}
		\begin{split}
			\oo'_{n-1} \rightarrow \frac{1}{n^2} |a'|^{-\frac{(n-2)^2}{2(n-1)}} :e^{+i \frac{n}{2}(\phi_5-\phi_6+\tilde{\phi}_5 - \tilde{\phi}_6)}: \,.
		\end{split}
		\label{eq:lift_leading_CPCPdown}
		\end{equation}
The normalized 3-point function of interest is
	\begin{equation}
	\begin{split}
		&\frac{\langle  \oo_n^\dagger(\infty,\infty) \oo^\dagger_2(1,1) \oo'_{n-1}(0,0) \rangle  }{\langle \oo^\dagger_n(\infty,\infty) \oo_n(0,0) \rangle }
		 \times \frac{\langle \oo^\dagger_n(1,1)\oo_n(0,0) \rangle^{1/2}}{\langle \oo^{\prime \dagger}_{n-1}(1,1) \oo'_{n-1}(0,0) \rangle^{1/2}\langle  \oo_2^\dagger(1,1)  \oo_2(0,0)\rangle^{1/2}} \,.
	\end{split}
	\label{leading_3pf_CPCP_down}
	\end{equation}
Then, the unnormalized twist-only part is computed using (\ref{eq:lift_leading_CPCPdown}), and (\ref{expOperatorForm}) with $q=-(n-1)/2$ and $q=-1/2$. The result is
	\begin{align}
		& \frac{1}{n^2} |a'_0|^{-\frac{(n-2)^2}{2(n-1)}} |a'_1|^{-\frac{1}{4}} |a'_\infty|^{-\frac{(n-1)^2}{2n}}
		\left| \langle :e^{i\frac{n}{2}(\phi_5-\phi_6)}(0)::e^{-\frac{i}{2}(\phi_5-\phi_6)}(1)::e^{-\frac{n-1}{2}(\phi_5-\phi_6)}(\infty): \rangle \right|^2 \nonumber\\
		&= \frac{1}{n^2} |a'_0|^{-\frac{(n-2)^2}{2(n-1)}} |a'_1|^{-\frac{1}{4}} |a'_\infty|^{-\frac{(n-1)^2}{2n}} |0-1|^{-n}|0-\infty|^{-n(n-1)}|1-\infty|^{n-1} \,.
	\end{align}
Since chiral primary operators are already normalized, the normalization here simply amounts to canceling divergent terms and flipping the sign in the exponent of $a'_\infty$. Adding the twist contribution (\ref{3ptBare}) with $n \rightarrow n-1$, the full result of (\ref{leading_3pf_CPCP_down}) becomes
	\begin{equation}
	\begin{split}
		 & \frac{1}{n^2} |a'_0|^{-\frac{(n-2)^2}{2(n-1)}} |a'_1|^{-\frac{1}{4}} |a'_\infty|^{+\frac{(n-1)^2}{2n}} \times 2^{-\frac{5}{4}}(n-1)^{-\frac{2n^2-n+2}{4n}}n^{\frac{2n^2-3n+3}{4(n-1)}}
= \frac{1}{2n(n-1)}
\label{leading_coefficient_CPCP_down}
	\end{split}
	\end{equation}
where we substituted the map coefficients (\ref{eq:twist_down_map_coeffs}). Squaring the result indeed reproduces the leading order coefficient in (\ref{eq:twist_down_4pf}).

Having worked out some higher order calculations explicitly in the previous section, here we only present the operators which account for the exchange channel. Through (\ref{eq:lift_leading_CPCPdown}) we find $\xi = n/(n-1)$ which fixes the basis (\ref{flowbasis}). The $w^{1/(n-1)}$ coefficient is accounted for by a single excitation on the leading operator
	\begin{equation}
		\sqrt{2}n^{-\frac{1}{n-1}} {J}^{\prime 3}_{-\frac{1}{n-1}}
		\ket{\oo'_{n-1}}\,.
	\end{equation}
Indeed, following the conjecture this is the only excitation without charge that we would expect to contribute. At the next order $w^{2/(n-1)}$ a combination of orthogonalized excitations act on the leading operator to account for the exchange channel
	\begin{align}
		&\frac{1}{6} n^{\frac{n+1}{1-n}} \bigg(
		(2 n+3) \left(3 {J}^{\prime 3}_{-\frac{2}{n-1}}+\sqrt{3} L'_{-\frac{2}{n-1}}\right)
		\nonumber\\
		& \quad + \sqrt{\frac{5}{3}} n \left(6 \left({J}^{\prime 3}_{-\frac{1}{n-1}}\right)^2 - (n-1) L'_{-\frac{2}{n-1}} \right)\bigg)
		\ket{\oo'_{n-1}}\,.
	\end{align}
At order $w^{3/(n-1)}$ the contributing excitations combine into
	\begin{align}
		\frac{n^{\frac{-2 n-1}{n-1}}}{2 \sqrt{3}}
		\bigg( &
		\left(2 n^2+7 n+4\right) \left(\sqrt{2} {J}^{\prime 3}_{-\frac{3}{n-1}}+{L'}_{-\frac{3}{n-1}}\right)
		\nonumber\\
		& \quad +\sqrt{5} (2 n+3) n \left({J}^{\prime 3}_{-\frac{2}{n-1}}{J}^{\prime 3}_{-\frac{1}{n-1}}-\frac{n-1}{12} {L'}_{-\frac{3}{n-1}} \right)
		\nonumber\\
		 & \quad + \sqrt{10} n^2 \left(\left({J}^{\prime 3}_{-\frac{1}{n-1}}\right)^3-\frac{3 (n-1)}{8} {J}^{\prime 3}_{-\frac{1}{n-1}}L'_{-\frac{2}{n-1}} \right)
		 \nonumber\\
		 & \quad +\sqrt{6}n(n+2) {J}^{\prime 3}_{-\frac{1}{n-1}}L'_{-\frac{2}{n-1}}
		 \bigg)\ket{\oo'_{n-1}} \,.
	\end{align}
We have also succeeded in matching the 4th and 5th order coefficients in (\ref{eq:twist_down_4pf}) perfectly. The rather lengthy details are relegated to Appendix {\ref{Appendix_B}.

\subsection{Fusing twist \texorpdfstring{$n$}{} CP and twist \texorpdfstring{$2$}{} anti CP to twist \texorpdfstring{$n+1$}{}}

We now turn our focus to investigating the OPE between a twist $n$ chiral primary and a twist $2$ {\em anti\,}-chiral primary. By flipping the roles of $t_+$ and $t_-$ one finds
	\begin{align}
		&\frac{\langle \oo^{\dagger}_{n}(\infty,\infty) \oo_{2}(1,1) \oo^{\dagger}_{2}(w,\bar{w}) \oo_{n}(0,0) \rangle}	{\langle \oo^{\dagger}_{n}(\infty,\infty)\oo_{n}(0,0) \rangle\langle \oo^{\dagger}_{2}(1,1) \oo_{2}(0,0) \rangle}
	=\frac{(n+1)^2}{4}n^{\frac{2-6n}{n+1}}|w|^{\frac{2-2n}{n+1}}\nonumber\\
		&\quad \times \bigg[ \bigg( 1+2(n-1)^2n^{-\frac{2n}{n+1}}w^{\frac{1}{n+1}} +n^{-\frac{4n}{n+1}}(3-14n+23n^2-14n^3+3n^4)w^{\frac{2}{n+1}}\nonumber\\
		&\quad \quad + \frac{2}{3} (n-1)^2 n^{-\frac{6 n}{n+1}} \left(6 n^4-35 n^3+67 n^2-35 n+6\right)w^{\frac{3}{n+1}}
	+\oo(w^{\frac{4}{n+1}})\bigg) \times\text{c.c.} \bigg] \,.
	\label{eq:CPn_ACP2_twist_up_4pf_coincident}
	\end{align}
The operator exchanged at leading order in the OPE must have twist $n+1$, and conformal weight and charge
	\begin{equation}
		h = \frac{n}{2}- \frac{n-1}{n+1}\ , \qquad q = \frac{n-2}{2} \,.
		\label{eq:weight_charge_leading_CPACP_up}
	\end{equation}
A natural operator with such properties is
	\begin{align}
		\oo_{NCP} \equiv \begin{cases}
		&J_{-\frac{n-3}{n+1}}^+\cdots J_{-\frac{1}{n+1}}^+\tilde{J}_{-\frac{n-3}{n+1}}^{\dot{+}}\cdots \tilde{J}_{-\frac{1}{n+1}}^{\dot{+}}\sigma_{n+1} \qquad n\text{ even}\\[8pt]
		&J_{-\frac{n-3}{n+1}}^+\cdots J_{-\frac{2}{n+1}}^+\tilde{J}_{-\frac{n-3}{n+1}}^{\dot{+}}\cdots \tilde{J}_{-\frac{2}{n+1}}^{\dot{+}}\sigma^{+\dot{+}}_{n+1} \qquad n\text{ 	odd}  \,.
	\end{cases}
	\label{eq:O_NCP}
	\end{align}
Interestingly, this is a not a chiral primary operator. It may be recognized as the chiral primary operator without the highest weight R-currents (c.f. (\ref{eq:chiral_primary})). To lift this to the cover, we use the Lunin-Mathur result that found
	\begin{equation}
		J_{-\frac{2k-1}{n+1}}^+\cdots J_{-\frac{1}{n+1}}^+\sigma_{n+1}\rightarrow a^{-\frac{k^2}{n+1}}e^{ik(\phi_5-\phi_6)} \,.
	\end{equation}
For $k=(n-2)/2$ one gets
	\begin{equation}
		\oo_{NCP}\rightarrow |a|^{-\frac{(n-2)^2}{2(n+1)}}e^{i\frac{n-2}{2}(\phi_5-\phi_6+\tilde{\phi}^5-\tilde{\phi}^6)} \,.
	\label{eq:lifted_O_NCP}
	\end{equation}
The relevant structure constant is computed by
	\begin{equation}
	\begin{split}
		&\frac{\langle \oo^{\dagger}_{n}(\infty,\infty) \oo_{2}(1,1) \oo_{NCP}(0,0) \rangle}{\langle \oo_n^\dagger(\infty,\infty)\oo_n(0,
	0) } \times
		\frac{\langle \oo_n^\dagger(1,1) \oo_n(0,0) \rangle^{1/2}}{\langle \oo_{NCP}^\dagger(1,1) \oo_{NCP}(0,0)\rangle^{1/2} \langle \oo_2^\dagger(1,1) \oo_2(0,0) \rangle^{1/2}} \,.
	\end{split}
	\label{eq:3pf_leading_CPACP_up}
	\end{equation}
Using (\ref{eq:lifted_O_NCP}) and (\ref{expOperatorForm}) with $q=-(n-1)/2$ and $q=1/2$ we find the unnormalized non-twist part
	\begin{equation}
	\begin{split}
		|a_0|^{-\frac{(n-2)^2}{2(n+1)}}|a_1|^{-1/4}|a_{\infty}|^{\frac{(n-1)^2}{2n}} |0-1|^{n+2}|0-\infty|^{-(n-2)(n-1)}|1-\infty|^{-(n-1)} \,.
	\end{split}
	\end{equation}
The full result of (\ref{eq:3pf_leading_CPACP_up}) is then obtained by normalizing and adding the twist contribution (\ref{3ptBare})
	\begin{equation}
	\begin{split}
		|a_0|^{-\frac{(n-2)^2}{2(n+1)}}|a_1|^{-1/4}|a_{\infty}|^{\frac{(n-1)^2}{2n}} |0-1|^{n+2} \times 
		2^{-\frac{5}{4}}(n+1)^\frac{2n^2+n+2}{4n}n^{-\frac{2n^2+3n+3}{4(n+1)}} = \frac{n+1}{2}n^{\frac{1-3n}{n+1}}
	\end{split}
	\end{equation}
the square of which is indeed the leading order contribution of the coincidence limit (\ref{eq:CPn_ACP2_twist_up_4pf_coincident}).

We continue with non-leading orders, following the previous section we provide the operators appearing in the OPE. The first non-leading coefficient is accounted for by
	\begin{equation}
		\sqrt{2}(n-1) n^{-\frac{n}{n+1}} {J'}^3_{-\frac{1}{n+1}} \ket{\oo_{NCP}} \,.
	\end{equation}
For the second non-leading coefficient, we find
	\begin{align}
		\frac{n^{-\frac{2n}{n+1}}}{18} \bigg( &
		\sqrt{3} \left(\sqrt{5} n^3-\left(\sqrt{5}+6\right) n^2-\left(\sqrt{5}-21\right) n+\sqrt{5}-6\right) L'_{-\frac{2}{n+1}}
		\nonumber\\ &  +9  \left(2 n^2-5 n+2\right) {J'}^3_{-\frac{2}{n+1}}  -6 \sqrt{15}  (n-1)^2 \left({J'}^3_{-\frac{1}{n+1}}\right)^2 \bigg)\ket{\oo_{NCP}}
	\end{align}
and for the third non-leading coefficient, we find
	\begin{align}
		&\frac{(n-1) n^{-\frac{3n}{n+1}}}{144}
		\bigg(24 \sqrt{30} \left({J'}^3_{-\frac{3}{n+1}}\right)^3 (n-1)^2
		\nonumber\\&
		-9 \sqrt{2} {J'}^3_{-\frac{1}{n+1}}L'_{-\frac{2}{n+1}} \left(\sqrt{15} n^3-\left(\sqrt{15}+8\right) n^2-\left(\sqrt{15}-32\right) n+\sqrt{15}-8\right)
		\nonumber\\&
		-2 \sqrt{3} \bigg(12 \sqrt{5} {J'}^3_{-\frac{2}{n+1}}{J'}^3_{-\frac{1}{n+1}} \left(2 n^2-5 n+2\right)-12 \sqrt{2} {J'}^3_{-\frac{3}{n+1}} \left(2 n^2-7 n+2\right)
		\nonumber\\&
		+{L'}_{-\frac{3}{n+1}} \left(-2 \sqrt{5} n^3+3 \left(\sqrt{5}+8\right) n^2+3 \left(\sqrt{5}-36\right) n-2 \sqrt{5}+24\right)\bigg)\bigg)
		\ket{\oo_{NCP}} \,.
	\end{align}

\subsection{Fusing twist \texorpdfstring{$n$}{} CP and twist \texorpdfstring{$2$}{} anti CP to twist \texorpdfstring{$n-1$}{}}

As a last variation, we consider the twist down coincidence limit of the previous section,
	\begin{align}
		& \frac{\langle  \oo_n^\dagger(\infty,\infty)  \oo_2(1,1) \oo_2^\dagger(w,\bar{w}) \oo_n(0,0)  \rangle}{\langle \oo_2(0,0) \oo_2^\dagger(1,1) \rangle  \langle \oo_n(0,0) \oo^\dagger_n(\infty,\infty)  \rangle}
		= \left( \frac{n}{2(n-1)}\right)^2 |w|^{\frac{2-2n}{n-1}}
		\nonumber\\ & \qquad \times \bigg[ \bigg( 1 + 2 n^{-\frac{2 n}{n-1}}w^{\frac{1}{n-1}} + n^{-\frac{4 n}{n-1}} \left(3 n^2+4 n+3\right)w^{\frac{2}{n-1}}
		\nonumber\\ & \qquad \quad + 2 n^{-\frac{6 n}{n-1}} \left(2 n^4+7 n^3+11 n^2+7 n+2\right)w^{\frac{3}{n-1}}
		+\oo(w^{\frac{4}{n+1}})
		) \times ( \text{c.c.} \bigg) \bigg]\,.
	\label{eq:CPnACP2_twist_down_4pf_coincident}
	\end{align}
From this, in the twist down channel the leading order exchanged operator in the $\mathcal{O}_n \mathcal{O}^\dagger_2$ OPE must have twist $n-1$, and conformal weight and charge $h=q=(n-2)/2$. One would (correctly) guess this operator to be the chiral primary $\mathcal{O}_{n-1}$. The associated structure constant we need to compute is given by
	\begin{equation}
	\begin{split}
		\frac{\langle \oo^\dagger_n(\infty,\infty)\oo_2(1,1)\oo_{n-1}(0,0) \rangle}{\langle \oo^\dagger_n(\infty,\infty)\oo_n(0,0)\rangle}
		\times \frac{\langle \oo^\dagger_n(0,0)\oo_n(1,1)\rangle^{1/2} }{\langle \oo^\dagger_{n-1}(1,1)\oo_{n-1}(0,0)\rangle^{1/2} \langle \oo^\dagger_2(1,1)\oo_2(0,0)\rangle^{1/2}} \,.
	\end{split}
	\end{equation}		
The unnormalized non-twist three point function is
		\begin{equation}
	\begin{split}
		 |a'_0|^{-\frac{(n-2)^2}{2(n-1)}} |a'_1|^{-\frac{1}{4}} |a'_\infty|^{-\frac{(n-1)^2}{2n}}  |0-1|^{n-2} |0-\infty|^{-(n-2)(n-1)} |1-\infty|^{-(n-1)} \,.
	\end{split}
	\end{equation}
Normalizing and adding the twist term, (\ref{3ptBare}) with $n \rightarrow n-1$ gives
		\begin{equation}
	\begin{split}
		 \frac{n}{2(n-1)}\,.
	\end{split}
	\end{equation}
This confirms that $\oo_{n-1}$ is the leading order exchanged operator in the twist down channel of the $\oo_n \oo^\dagger_2$ OPE. It is not surprising that this result differs by a factor $n^2$ from  (\ref{leading_coefficient_CPCP_down}). At higher orders, the switching of daggers will affect the dressing terms.

Lastly, we quote the operators that appear in the OPE and account for the first three non-leading orders. At first order we find
	\begin{equation}
		-\sqrt{2} n^{-\frac{n}{n-1}}{J'}^3_{-\frac{1}{n-1}}\ket{\oo_{n-1}} \,.
	\end{equation}
At second order the result is
	\begin{align}
		\frac{1}{18} n^{-\frac{2 n}{n-1}} &\bigg(6 \sqrt{15} {J'}^3_{-\frac{1}{n-1}}{J'}^3_{-\frac{1}{n-1}}-9  (3 n+2) {J'}^3_{-\frac{2}{n-1}}
		\nonumber\\ &+\sqrt{3}  \left(n(9-\sqrt{5}) +6+\sqrt{5}\right){L'}_{-\frac{2}{n-1}} \bigg)\ket{\oo_{n-1}} \,.
	\end{align}
Lastly, the third order result is
	\begin{align}
		&\frac{1}{144} n^{-\frac{3 n}{n-1}} \bigg(-24 \sqrt{30} \left( {J'}^3_{-\frac{1}{n-1}}\right)^3+9 \sqrt{2} \left(\left(\sqrt{15}-16\right) n-\sqrt{15}-8\right) {J'}^3_{-\frac{1}{n-1}}L'_{-\frac{2}{n-1}}
		\nonumber\\&\quad+2 \sqrt{3} \bigg(12 \sqrt{5}(3 n+2)  {J'}^3_{-\frac{2}{n-1}} {J'}^3_{-\frac{1}{n-1}} -12 \sqrt{2} \left(4 n^2+7 n+2\right) {J'}^3_{-\frac{3}{n-1}}
		\nonumber\\&\quad+\left(-3 \left(\sqrt{5}-16\right) n^2+\left(\sqrt{5}+84\right) n+2 \left(\sqrt{5}+12\right)\right) {L'}_{-\frac{3}{n-1}} \bigg)\bigg)
		\ket{\oo_{n-1}} \,.
	\end{align}

\section{Discussion}\label{discussion}
In this work, we have extended our analysis of \cite{Burrington:2018upk} to the $\mathcal{N}=(4,4)$ supersymmetric case, and likewise extended our conjecture. Specifically, we conjecture that the operators connected to bare twists by spectral flow parallel to the twist, along with their excitations using the fractional modes of the current algebra, form a closed subalgebra of the operator algebra at large $N$.  The evidence for this conjecture is as follows.  The (anti) chiral primaries and all operators related to them by spectral flow parallel to the twist have 4-point functions that are universal at large $N$.  This directly leads to the universality of the 4-point functions of the operators constructed using the fractional modes of the $\mathcal{N}=(4,4)$ superconformal algebra acting on these base operators, simply because they are algebraically constructed.  Therefore, the crossing channels are universal as well.  This self consistency means that this subspace of operators at least has the potential to be closed under the OPE.  As direct evidence, we have considered some specific 4-point functions.  We have shown that the exchange channels are indeed exactly reproduced by the operators constructed using the fractional modes of the $\mathcal{N}=(4,4)$ algebra, and so we have shown that the operator algebra is closed, at least to the several non-trivial orders to which we expand.  We conjecture that this is true at all orders, and for all operators constructed in this way.

Part of our motivation for investigating OPEs between twisted operators in symmetric orbifold SCFTs is to enable exploration of the deformation away from the free orbifold point of the D1-D5 CFT. Indeed, the deformation operator itself is a twist 2 operator, and this was our main motivation to consider OPEs between operators in the twist $n$ sector and the twist $2$ sector. Specifically, the deformation operator is a twist 2 operator dressed with SUSY currents,
	\begin{equation}
		\mathcal{O}_D = \epsilon_{\alpha\beta} \epsilon_{\dot{\alpha}\dot{\beta}} \epsilon_{AB} G_{-1/2}^{\alpha A} \tilde{G}_{-1/2}^{\dot{\alpha}B} 	\sigma_2^{\beta\dot{\beta}}\,.
	\end{equation}
On the cover it lifts to
	\begin{equation}
		\mathcal{O}_D \rightarrow -i|b|^{-5/4} \epsilon_{AB} :\partial X_{\dot{A}A} \bar{\partial}X_{\dot{B}B} \mathcal{S}^{\dot{A}\dot{B}}: \label{defopform}
	\end{equation}
where the ${\mathcal{S}}S^{\dot{A}\dot{B}}$ are the spin fields on the cover transforming under the appropriate $SU(2)$ symmetries.  Note that this operator is in the class of operators discussed in the previous paragraph.  When deforming away from the orbifold point, the mixing of two operators $O_{1}$ and $O_2$ is governed by the three point functions
\begin{equation}
\langle {O}_2 {\mathcal O}_D {O}_1\rangle\,.
\end{equation}
Understanding the structure of the OPE directly would make these calculations much more tractable, furnishing the operators ${O}_2$ directly, which is the ultimate goal of our explorations here and in \cite{Burrington:2018upk,Burrington:2017jhh}.  We plan to further explore this in future work.

Note that the repeated indices in (\ref{defopform}) are summed over, such that the deformation operator is a sum of operators with different charges under the various $SU(2)$s.  Therefore, when working with bosonized fermions, the cocycles become important to properly track relative phases \cite{Burrington:2015mfa}. Since we were working directly with pure exponentials, we were able to avoid this complication here, as explained in Section \ref{section2}.  While the addition of cocycles makes computations more cumbersome, it does not add any fundamental challenge.

We have begun to study the structure of the OPE between a twist $n$ operator and the deformation operator.  To do so, we could consider the following four point function of the form	
	\begin{equation}
		\langle  \oo^\dagger_n(\infty,\infty) \oo_D(1,1) \oo_D(w,\bar{w})   \oo_n(0,0) \rangle
	\end{equation}
where we are omitting the dagger on $\oo_D$ since this operator is real. The four point function above including normalization may be obtained using covering space techniques,
	\begin{align}
		&\frac{\langle \mathcal{O}^{\dagger}_{n}(\infty,\infty) \mathcal{O}_{D}(1,1) \mathcal{O}_{D}(w,\bar{w}) \mathcal{O}_{n}(0,0) \rangle}{\langle \mathcal{O}^{\dagger}_{n}(\infty,\infty) \mathcal{O}_{n}(0,0) \rangle\langle \mathcal{O}_{D}(1,1) \mathcal{O}_{D}(0,0)\rangle}\nonumber\\
		&=|C|^{-4}n^2|a|^{2(1-n)}|1-a|^{-2}|(n-1)^2a-(n+1)^2|^{-4}\,.
	\end{align}
Expanded around $a = 0$, i.e., for the fusion $\sigma_n \sigma_2 \rightarrow \sigma_{n+1}$, the result becomes
	\begin{align}
	&\frac{\langle \mathcal{O}^{\dagger}_{n}(\infty,\infty)\mathcal{O}_{D}(1,1) \mathcal{O}_{D}(w,\bar{w}) \mathcal{O}_{n}(0,0) \rangle}{\langle\mathcal{O}^{\dagger}_{n}(\infty,\infty)\mathcal{O}_{n}(0,0)\rangle\langle\mathcal{O}_{D}(1,1)\mathcal{O}_{D}(0,0)\rangle}\nonumber\\
	&=n^{\frac{2-6n}{n+1}}|w|^{\frac{2-2n}{n+1}}\bigg[\bigg(1+3(n-1)^2n^{-\frac{2n}{n+1}}w^{\frac{1}{n+1}}\nonumber\\
	&+2n^{-\frac{4n}{n+1}}(3-14n+23n^2-14n^3+3n^4)w^{\frac{2}{n+1}}+\mathcal{O}(w^{\frac{3}{n+1}})\bigg)\times\text{c.c.}\bigg] \,.
	\end{align}
This indicates that the leading order exchanged operator has charge $\frac{n-1}{2}$ and weight $\frac{n^2+3}{2(n+1)}$. This looks similar to equation (\ref{eq:weight_charge_leading_CPACP_up}), which suggests an operator like
\begin{align}\label{non-chiralprim2a}
&\epsilon_{AB}G^{+A}_{-1/2}J_{-\frac{n-3}{n+1}}^+\cdots J_{-\frac{1}{n+1}}^+\tilde{G}^{\dot{+}B}_{-1/2}\tilde{J}_{-\frac{n-3}{n+1}}^{\dot{+}}\cdots \tilde{J}_{-\frac{1}{n+1}}^{\dot{+}}\sigma_{n+1} \qquad n\text{ even}\\[8pt]\label{non-chiralprim2b}
&\epsilon_{AB}G^{+A}_{-1/2}J_{-\frac{n-3}{n+1}}^+\cdots J_{-\frac{2}{n+1}}^+\tilde{G}^{\dot{+}B}_{-1/2}\tilde{J}_{-\frac{n-3}{n+1}}^{\dot{+}}\cdots \tilde{J}_{-\frac{2}{n+1}}^{\dot{+}}\sigma^{+\dot{+}}_{n+1} \qquad n\text{ odd} \,.
\end{align}
Such a guess would not only make sense given the application of $G_{-1/2}$ to the twist to chiral primary to get the deformation, but also has the correct weight and charge. We look forward to undertaking a more systematic exploration of deformation physics in the D1-D5 SCFT.

There are other interesting future avenues of investigation as well.  In this paper we have limited ourselves to $\mathcal{N}=(4,4)$ supersymmetric CFTs. The choice to work in the $c=6$ representation of the $\mathcal{N}=(4,4)$ SUSY algebra connected to the D1-D5 system involved no loss of generality, as explained in the introduction. Altering the symmetry algebra, on the other hand, produces important differences, although we believe that in cases with extended supersymmetry similar patterns should emerge. Thus, it is natural to test our conjecture in theories with $\mathcal{N}=(3,3)$ and ${\mathcal{N}}=(2,2)$ because the spectrum would still offer BPS states similar to those discussed here, presumably also related by spectral flow parallel to twists.  This is in contrast to the case $\mathcal{N}=(1,1)$. One way to obtain an $\mathcal{N}=(3,3)$ algebra is to consider the orbifold $\mathcal{S}_0/\mathds{Z}_2$, where $\mathcal{S}_0$ is a member of the $\mathcal{N}=(4,4)$ family $\mathcal{S}_\kappa$. Recently it was proposed in \cite{Eberhardt:2018sce} that string theory on AdS$_3 \times (S^3 \times S^3 \times S^1)/\mathds{Z}_2$ is dual to the supersymmetric product orbifold $(S_0/\mathds{Z}_2)^N/S_N$, which has central charge $c=3N$. There are four different sectors, as one has to account for even or odd parity under $\mathds{Z}_2$ and even or odd $n$ twist for the $S_N$. We have begun work in this direction and anticipate that fleshing out the story for less supersymmetric cases will prove enlightening.

\section*{Acknowledgements}
The work of AWP and TdB is supported by a Discovery Grant from the Natural Sciences and Engineering Research Council of Canada.  The work of BAB is supported by grants from Hofstra University. BAB is thankful for support from the Scholars program at KITP, which is supported in part by the National Science Foundation under Grant No. NSF PHY-1748958.

\appendix

\section{Fractional algebra sample calculations}\label{Appendix_A}

Here, we wish to find the commutation relations between fractional modes of currents in the presence of a (possibly excited) $\sigma_{(123\cdots n)}'=\sigma_{n}'$ cycle twist field.  To do so, we will use covering space techniques \cite{Lunin:2000yv,Lunin:2001pw,Burrington:2012yn}.  To compute the action of a fractional mode of a current $Q$ on a state in the twist $n$ sector $\sigma_n'$, one lifts to the covering surface via a map $z(t)$, resulting in a contour integral on the surface of the form
\begin{equation}
Q^{\frac{\ell}{n}}\ket{ \sigma_n'} \rightarrow \oint \frac{dt}{2\pi i} \pa z(t) (z(t))^{h_Q-1+\frac{\ell}{n}}Q^\uparrow(t) (\sigma_n')^\uparrow(0)
\end{equation}
where $Q^\uparrow(t)$ is the conformally mapped current, and $(\sigma_n')^\uparrow(0)$ is the operator representation of the twist $n$ sector state $\ket{ \sigma_n'}$ lifted to the cover (above we have assumed that the ramified point in the map on the cover is at $t=0$ for convenience).  One must be careful to include the non-tensorial transformation properties of the current when $Q=T$.  Using the above, one can compute the commutation relations of the fractional modes by simply evaluating the contour integrals on the cover \cite{Burrington:2018upk,Burrington:2012yn,Roumpedakis:2018tdb}.

We give two of the more interesting calculations as examples, and simply state the results for other calculations.  We consider the $T\cdot J$ commutation relations, and find
\begin{align}
\left[L_{\frac{\ell}{n}},J^{i}_{\frac{\ell'}{n}}\right]\ket{ \sigma'} & \rightarrow  \oint \frac{dt_2}{2\pi i} \oint_{t_1=t_2} \frac{dt_1}{2\pi i} \frac{z(t_1)^{\frac{\ell}{n}+1}}{\partial z(t_1)} z(t_2)^\frac{\ell'}{n} \left(T(t_1)-\frac{c}{12} \left\{z(t_1),t_1\right\}\right) J^i(t_2) \sigma'_\uparrow\,.
\end{align}
As in the $T\cdot T$ case \cite{Burrington:2018upk}, the Schwarzian is non-singular, since $t_2$ is a non-ramified point and so only the $T\cdot J$ OPE contributes.  Using this, we find
\begin{align}
&  \oint \frac{dt_2}{2\pi i} \oint_{t_1=t_2} \frac{dt_1}{2\pi i} \frac{z(t_1)^{\frac{\ell}{n}+1}}{\partial z(t_1)} z(t_2)^\frac{\ell'}{n} \left(\frac{J^i(t_2)}{t_{12}^2}+ \frac{\partial J^i(t_2)}{t_{12}}\right)\sigma'_\uparrow\,.
\end{align}
We expand $\frac{z(t_1)^{\frac{\ell}{n}+1}}{\partial z(t_1)}$ in $t_{12}$ to find
\begin{equation}
\frac{z(t_1)^{\frac{\ell}{n}+1}}{\partial z(t_1)}=\frac{z(t_2)^{\frac{\ell}{n}+1}}{\partial z(t_2)} + t_{12} \left(\left(\frac{\ell}{n}+1\right)z(t_2)^\frac{\ell}{n}-z(t_2)^{\frac{\ell}{n}+1} \frac{\partial^2 z(t_2)}{(\partial z(t_2))^2}\right)+\cdots \label{Tzt1expand}
\end{equation}
allowing us to perform the $t_1$ contour integral, giving
\begin{align}
& \left[L_{\frac{\ell}{n}},J^{i}_{\frac{\ell'}{n}}\right]\ket{ \sigma'} \nonumber \\
&\rightarrow
\oint \frac{dt_2}{2\pi i} \left[ \frac{z(t_2)^{\frac{\ell+\ell'}{n}+1}}{\partial z(t_2)} \partial J^i(t_2)+\left(\frac{\ell}{n}+1\right)z(t_2)^\frac{\ell+\ell'}{n}J^i(t_2)-z(t_2)^{\frac{\ell+\ell'}{n}+1} \frac{\partial^2 z(t_2)}{\left(\partial z(t_2)\right)^2}J^i(t_2)\right]\sigma'_\uparrow \,.
\end{align}
The first term we integrate by parts, finding
\begin{equation}
\left[L_{\frac{\ell}{n}},J^{i}_{\frac{\ell'}{n}}\right]\ket{ \sigma'} \rightarrow
\oint \frac{dt_2}{2\pi i} \left(-\frac{\ell'}{n}\right)z(t_2)^\frac{\ell+\ell'}{n}J^i(t_2)\sigma'_\uparrow
\end{equation}
which we recognize as a fractional mode of the $R$ current acting on the state $\ket{\sigma'_n}$.  Thus, we conclude that
\begin{equation}
\left[L_{\frac{\ell}{n}},J^{i}_{\frac{\ell'}{n}}\right]=-\frac{\ell'}{n} J^i_{\frac{\ell'+\ell}{n}}
\end{equation}
giving the straightforward generalization of the algebra to fractional modes.

\newpage
Let us also show how the $G\cdot\hat{G}$ calculation works.
\begin{align}
\left\{ G^{\alpha}_{\frac{\ell}{n}}, \hat{G}_{\beta,\frac{\ell'}{n}}\right\}\ket{ \sigma'} & \rightarrow \oint \frac{dt_2}{2\pi i} \oint_{t_1=t_2} \frac{dt_1}{2\pi i} \frac{z(t_1)^{\frac{\ell}{n}+\frac{1}{2}}}{\left(\partial z(t_1)\right)^\frac{1}{2}} \frac{z(t_2)^{\frac{\ell'}{n}+\frac{1}{2}}}{\left(\partial z(t_2)\right)^\frac{1}{2}} G^\alpha(t_1) \hat{G}_{\beta}(t_2)\sigma'_\uparrow\nonumber \\
& = \oint \frac{dt_2}{2\pi i} \oint_{t_1=t_2} \frac{dt_1}{2\pi i} \frac{z(t_1)^{\frac{\ell}{n}+\frac{1}{2}}}{\left(\partial z(t_1)\right)^\frac{1}{2}} \frac{z(t_2)^{\frac{\ell'}{n}+\frac{1}{2}}}{\left(\partial z(t_2)\right)^\frac{1}{2}} \nonumber \\
&
\qquad\times \left(\frac{2}{3}\frac{c \delta^\alpha_{\ \beta}}{t_{12}^3}+4\frac{(\sigma^{i*})^\alpha_{\ \beta} J^i(t_2)}{t_{12}^2}+\frac{2T(t_2)\delta^\alpha_{\ \beta} + 2 (\sigma^i)^\alpha_{\ \beta} \partial J^i(t_2)}{t_{12}}\right)\sigma'_\uparrow\,.
\end{align}
We expand
\begin{align}
\frac{z(t_1)^{\frac{\ell}{n}+\frac{1}{2}}}{\left(\partial z(t_1)\right)^\frac{1}{2}} & =\frac{z(t_2)^{\frac{\ell}{n}+\frac{1}{2}}}{\left(\partial z(t_2)\right)^\frac{1}{2}}+ t_{12}\left(\left(\frac{\ell}{n}+\frac{1}{2}\right) z(t_2)^{\frac{\ell}{n}-\frac{1}{2}}\left(\partial z(t_2)\right)^\frac{1}{2} -\frac{1}{2}\frac{z(t_2)^{\frac{\ell}{n}+\frac{1}{2}} \partial^2 z(t_2)}{\left(\partial z(t_2)\right)^{\frac{3}{2}}}\right) \nonumber \\
& \qquad +\frac{t_{12}^2}{2} \Bigg(\left(\frac{\ell}{n}+\frac{1}{2}\right)\left(\frac{\ell}{n}-\frac{1}{2}\right) z(t_2)^{\frac{\ell}{n}-\frac{3}{2}} \left(\partial z(t_2)\right)^\frac{3}{2} \nonumber\\
& \qquad \qquad \qquad +\frac{3}{4} z(t_2)^{\frac{\ell}{n}+\frac{1}{2}} \frac{\left(\partial^2 z(t_2)\right)^2}{\left(\partial z(t_2)\right)^{\frac{5}{2}}}-\frac{1}{2} \frac{z(t_2)^{\frac{\ell}{n}+\frac{1}{2}}\partial^3 z(t_2)}{\left(\partial z(t_2)\right)^\frac{3}{2}}\Bigg) + \cdots \,.
\end{align}
The last two terms in the $t_{12}^2$ part of the above expansion may be grouped into a Schwarzian derivative, and so
\begin{align}
\frac{z(t_1)^{\frac{\ell}{n}+\frac{1}{2}}}{\left(\partial z(t_1)\right)^\frac{1}{2}} & =\frac{z(t_2)^{\frac{\ell}{n}+\frac{1}{2}}}{\left(\partial z(t_2)\right)^\frac{1}{2}}+ t_{12}\left(\left(\frac{\ell}{n}+\frac{1}{2}\right) z(t_2)^{\frac{\ell}{n}-\frac{1}{2}}\left(\partial z(t_2)\right)^\frac{1}{2} -\frac{1}{2}\frac{z(t_2)^{\frac{\ell}{n}+\frac{1}{2}} \partial^2 z(t_2)}{\left(\partial z(t_2)\right)^{\frac{3}{2}}}\right) \nonumber \\
&+\frac{t_{12}^2}{2} \Bigg(\left(\frac{\ell}{n}+\frac{1}{2}\right)\left(\frac{\ell}{n}-\frac{1}{2}\right) z(t_2)^{\frac{\ell}{n}-\frac{3}{2}} \left(\partial z(t_2)\right)^\frac{3}{2} -\frac{z(t_2)^{\frac{\ell}{n}+\frac{1}{2}}}{2\left(\partial z(t_2)\right)^\frac{1}{2}} \left\{ z(t_2),t_2\right\}\Bigg) + \cdots \,.
\end{align}
\vskip-0.15truein
\noindent Using this, we find
\begin{align}
& \left\{ G^{\alpha}_{\frac{\ell}{n}}, \hat{G}_{\beta,\frac{\ell'}{n}}\right\}\ket{ \sigma'}
\nonumber \\
& \rightarrow \oint \frac{dt_2}{2\pi i} \Bigg( \frac{z(t_2)^{\frac{\ell+\ell'}{n}+1}}{\partial z(t_2)} 2 T(z_2) \delta^\alpha_{\ \beta} +\frac{z(t_2)^{\frac{\ell+\ell'}{n}+1}}{\partial z(t_2)}2(\sigma^{*i})^\alpha_{\ \beta} \partial J^i \nonumber \\ & \qquad +\left(\left(\frac{\ell}{n}+\frac{1}{2}\right)z(t_2)^\frac{\ell+\ell'}{n}- \frac{1}{2}\frac{z(t_2)^{\frac{\ell+\ell'}{n}+1}\partial^2z(t_2)}{\left(\partial z(t_2)\right)^2}\right)4(\sigma^{i*})^\alpha_{\ \beta} J^i(t_2) \nonumber\\
& \qquad \qquad +\frac{1}{3} c \delta^\alpha_{\ \beta} \left(\left(\frac{\ell}{n}+\frac{1}{2}\right)\left(\frac{\ell}{2}-\frac{1}{2}\right) z(t_2)^{\frac{\ell+\ell'}{n}-1}\partial z(t_2)-\frac{z(t_2)^{\frac{\ell+\ell'}{n}+1}}{2\partial_z} \left\{z(t_2),t_2\right\}\right)\Bigg)\sigma'_\uparrow\,.
\end{align}
\vskip-0.15truein
\noindent
The $\partial J$ term we integrate by parts, finally finding
\begin{align}
& \left\{ G^{\alpha}_{\frac{\ell}{n}}, \hat{G}_{\beta,\frac{\ell'}{n}}\right\}\ket{ \sigma'} \nonumber \\
& \rightarrow \oint \frac{dt_2}{2\pi i} \Bigg(2\frac{z(t_2)^{\frac{\ell+\ell'}{n}+1}}{\partial z(t_2)} \left(T(t_2)-\frac{c}{12}\left\{z(t_2),t_2\right\}\right)\delta^\alpha_{\ \beta}+ 2\left(\frac{\ell}{n}-\frac{\ell'}{n}\right)(\sigma^{i*})^\alpha_{\ \beta} z(t_2)^{\frac{\ell+\ell'}{n}} J^i(t_2) \nonumber \\
& \qquad \qquad +\frac{1}{3} c \delta^{\alpha}_{\ \beta} \left(\frac{\ell}{n}+\frac{1}{2}\right)\left(\frac{\ell}{n}-\frac{1}{2}\right) z(t_2)^{\frac{\ell+\ell'}{n}-1}\partial z(t_2)\Bigg) \,.
\end{align}
Again, the last term above is a total derivative unless $\frac{\ell}{n}+\frac{\ell'}{n}=0$.  Using $z(t_2)=a t_2^n+\cdots$, and reading off the other terms as modes of the currents, we find
\begin{equation}
\left\{ G^{\alpha}_{\frac{\ell}{n}}, \hat{G}_{\beta,\frac{\ell'}{n}}\right\}=2\delta^{\alpha}_{\ \beta} L_{\frac{\ell+\ell'}{n}}+2 \left(\frac{\ell}{n}-\frac{\ell'}{n}\right)(\sigma^{i*})^\alpha_{\ \beta} J^i_{\frac{\ell+\ell'}{n}} + \frac{nc}{3} \left(\frac{\ell}{n}+\frac{1}{2}\right)\left(\frac{\ell}{n}-\frac{1}{2}\right)\delta^\alpha_{\ \beta} \delta_{\frac{\ell+\ell'}{n},0}
\end{equation}
again, giving the correct algebra with the central charge term $nc$ accounting for the $n$ copies parallel to the twist.

Completing the other four calculations associated with (\ref{JJOPE})-(\ref{JGhatOPE}) is straightforward, giving the algebra quoted in the text (\ref{LLcommute})-(\ref{GGcommute}).

\section{Higher order results}
\label{Appendix_B}
Here we provide results of matching 3-point function calculations to the 4th and 5th orders of both the twist up coincidence limit (\ref{CPCPTwistUpExpand}) and the twist down coincidence limit (\ref{eq:twist_down_4pf}).

At 4th order, the operators that can contribute are
	\begin{align}
		O^{(4)}_{1,n} &= \left( {J'}^3_{-\frac{1}{n}} \right)^4 O_n
		& \langle |O^{(4)}_{1,n}|^2 \rangle &=   \frac{3}{2}
		& O^{(4)}_{6,n} &=  \left({L'}_{-\frac{2}{n}}\right)^2 O_n
		& \langle |O^{(5)}_{6,n}|^2 \rangle &=  \frac{42}{n^4}	
		\nonumber\\
		O^{(4)}_{2,n} &= {J'}^3_{-\frac{2}{n}} \left( {J'}^3_{-\frac{1}{n}} \right)^2 O_n
		& \langle |O^{(4)}_{2,n}|^2 \rangle &=   \frac{1}{2}
		& O^{(4)}_{7,n} &=  {L'}_{-\frac{4}{n}}  O_n
		& \langle |O^{(5)}_{7,n}|^2 \rangle &=  \frac{30}{n^2}
		\nonumber\\
		O^{(4)}_{3,n} &= {J'}^3_{-\frac{3}{n}}  {J'}^3_{-\frac{1}{n}} O_n
		& \langle |O^{(4)}_{3,n}|^2 \rangle &=   \frac{3}{4}
		& O^{(4)}_{8,n} &=  \left( {J'}^3_{-\frac{1}{n}}\right)^2 {L'}_{-\frac{2}{n}} O_n
		& \langle |O^{(5)}_{8,n}|^2 \rangle &=  \frac{11}{4n^2}
		\nonumber\\
		O^{(4)}_{4,n} &= \left( {J'}^3_{-\frac{2}{n}} \right)^2 O_n
		& \langle |O^{(4)}_{4,n}|^2 \rangle &=   2
		& O^{(4)}_{9,n} &=  {J'}^3_{-\frac{2}{n}} {L'}_{-\frac{2}{n}} O_n
		& \langle |O^{(5)}_{9,n}|^2 \rangle &=  \frac{3}{n^2}
		\nonumber\\
		O^{(4)}_{5,n} &= {J'}^3_{-\frac{4}{n}} O_n
		& \langle |O^{(4)}_{5,n}|^2 \rangle &=   2
		& O^{(4)}_{10,n} &=  {J'}^3_{-\frac{1}{n}} {L'}_{-\frac{3}{n}} O_n
		& \langle |O^{(5)}_{10,n}|^2 \rangle &=  \frac{7}{n^2} \,.
	\end{align}
	For twist up calculations we let $n \rightarrow n+1$, while for twist down calculations $n \rightarrow n-1$. Below we provide the 3-point function results. While not explicitly written down on the left-hand sides, these 3-point functions are fully normalized, including the extra normalization due to the excitation operator, which may be an orthogonal combination of the operators above. The extra normalization was factored off explicitly; shown are the square root parts.

	In the twist up case, the 3-point functions at level $k=4$ are
	\begin{align}
		& \langle O^{(4)}_{7,n+1}       \mathcal{O}^\dagger_2 \mathcal{O}^\dagger_n \rangle
		= \frac{5 (-n)^{\frac{n+5}{n+1}} \left(3 n^2-7 n+3\right)}{2 n^4 (n+1)} \sqrt{\frac{(n+1)^2}{30}}C_{n,2,n+1}
		\nonumber\\
		& \langle O^{(4)}_{5,n+1}       \mathcal{O}^\dagger_2 \mathcal{O}^\dagger_n \rangle
		= \frac{(-n)^{\frac{n+5}{n+1}} \left(3 n^2-7 n+3\right)}{2 n^4} \sqrt{\frac{1}{2}}C_{n,2,n+1}
		\nonumber\\
		& \langle O^{(4)}_{9,n+1}       \mathcal{O}^\dagger_2 \mathcal{O}^\dagger_n \rangle
		= \frac{3 (-n)^{\frac{2-2 n}{n+1}}}{4 (n+1)} \sqrt{\frac{(n+1)^2}{3}}C_{n,2,n+1}
		\nonumber\\
		& \langle \left( O^{(4)}_{9,n+1} - \frac{n+1}{15} O^{(4)}_{7,n+1} \right)     \mathcal{O}^\dagger_2 \mathcal{O}^\dagger_n \rangle
		= \frac{1}{12} (-n)^{\frac{2-2 n}{n+1}} \sqrt{\frac{15}{28}}C_{n,2,n+1}
		\nonumber\\
		& \frac{1}{14} \langle \left( \frac{9}{(n+1)^2}O^{(4)}_{4,n+1}+14 O^{(4)}_{6,n+1}-\frac{9}{n+1}O^{(4)}_{7,n+1} \right)     \mathcal{O}^\dagger_2 \mathcal{O}^\dagger_n \rangle
		\nonumber\\
		&\qquad\qquad\qquad\qquad\qquad\qquad\quad\
		  = \frac{213 (-n)^{\frac{2-2 n}{n+1}}}{56 (n+1)^2} \sqrt{\frac{7(n+1)^4}{213}}C_{n,2,n+1} \,.
		  \label{eq:3pfs_k=4_up_firstlast}
	\end{align}
Recall that we had defined $C_{n,2,n+1} = \frac{n+1}{2n}$. Adding up the squares of (\ref{eq:3pfs_k=4_up_firstlast}) gives
	\begin{equation}
		(-n)^{\frac{2-6 n}{n+1}} \left(3 n^4-14 n^3+23 n^2-14 n+3\right)C_{n+1,2,n}^2 \,.
	\end{equation}
This correctly matches the order $w^{4/(n+1)}$ coefficient in the twist up coincidence limit (\ref{CPCPTwistUpExpand}).

The results for twist down 3-point functions at level $k=4$ are
	\begin{align}
		\langle O^{(4)}_{7,n-1} \mathcal{O}^\dagger_2 \mathcal{O}^\dagger_n \rangle
		&= \frac{n^{-\frac{1+3n}{n-1}} \left(6 n^3+35 n^2+47 n+15\right)}{2 (n-1)} \sqrt{\frac{(n-1)^2}{30}}C^{\star}_{n,2,n-1}
\nonumber\\
		\langle O^{(4)}_{5,n-1} \mathcal{O}^\dagger_2 \mathcal{O}^\dagger_n \rangle
		&= \frac{1}{6} n^{-\frac{1+3n}{n-1}} \left(6 n^3+35 n^2+47 n+15\right) \sqrt{\frac{1}{2}}C^{\star}_{n,2,n-1}
\nonumber\\
		\langle O^{(4)}_{9,n-1}
		 \mathcal{O}^\dagger_2 \mathcal{O}^\dagger_n \rangle
		&= \frac{n^{-\frac{2 (n+1)}{n-1}} \left(12 n^2+44 n+35\right)}{12 (n-1)} \sqrt{\frac{(n-1)^2}{3}}C^{\star}_{n,2,n-1}\nonumber
	\end{align}
	\vskip-0.3truein
	\noindent
	\begin{align}
	\langle \left( O^{(4)}_{2,n-1}-\frac{ n-1}{6}O^{(4)}_{9,n-1} \right)  \mathcal{O}^\dagger_2 \mathcal{O}^\dagger_n \rangle
	&= \frac{5}{36} n^{-\frac{2 (n+1)}{n-1}} \left(6 n^2+13 n+4\right) \sqrt{\frac{12}{5}}C^{\star}_{n,2,n-1}
\nonumber\\
		\langle \left( O^{(4)}_{4,n-1} - \frac{n-1}{15}O^{(4)}_{7,n-1} \right)
		 \mathcal{O}^\dagger_2 \mathcal{O}^\dagger_n \rangle
		&= \frac{1}{60} n^{-\frac{2 (n+1)}{n-1}} \left(48 n^2+140 n+101\right) \sqrt{\frac{15}{28}}C^{\star}_{n,2,n-1}
\nonumber\\
		\langle \left( O^{(4)}_{10,n-1}-\frac{2 }{n-1}O^{(4)}_{2,n-1} \right)
		 \mathcal{O}^\dagger_2 \mathcal{O}^\dagger_n \rangle
		&=\frac{10 n^{-\frac{2 (n+1)}{n-1}} (n+1)}{3 (n-1)} \sqrt{\frac{(n-1)^2}{5}}C^{\star}_{n,2,n-1}\nonumber
	\end{align}	
	\vskip-0.3truein
	\noindent
	\begin{align}
		&\langle \left( O^{(4)}_{6,n-1} + \frac{9}{14 (n-1)^2}O^{(4)}_{4,n-1}  - \frac{9 }{14 (n-1)}O^{(4)}_{7,n-1}  \right)
		 \mathcal{O}^\dagger_2 \mathcal{O}^\dagger_n \rangle
		\nonumber\\ &= \frac{n^{-\frac{2 (n+1)}{n-1}} \left(96 n^2+364 n+293\right)}{56 (n-1)^2} \sqrt{\frac{7(n-1)^4}{213}}C^{\star}_{n,2,n-1}
\nonumber\\
		\frac{1}{284} &\langle \left(  284 O^{(4)}_{1,n-1}-9 O^{(4)}_{4,n-1}+(n-1) (9 O^{(4)}_{7,n-1}-14 O^{(4)}_{6,n-1} (n-1))  \right)
		 \mathcal{O}^\dagger_2 \mathcal{O}^\dagger_n \rangle
		\nonumber\\ &= \frac{5}{284} n^{-\frac{2 (n+1)}{n-1}} \left(52 n^2+67 n-4\right) \sqrt{\frac{284}{405}}C^{\star}_{n,2,n-1}
\nonumber\\
		\frac{1}{216} &\langle \left( 4 O^{(4)}_{1,n-1}+216 O^{(4)}_{3,n-1}+9 O^{(4)}_{4,n-1}-4 O^{(4)}_{6,n-1} (n-1)^2-9 O^{(4)}_{7,n-1} (n-1) \right)
		 \mathcal{O}^\dagger_2 \mathcal{O}^\dagger_n \rangle
		 \nonumber\\ & =\frac{5}{216} n^{-\frac{2 (n+1)}{n-1}} \left(38 n^2+131 n+76\right) \sqrt{\frac{144}{95}}C^{\star}_{n,2,n-1}
\nonumber\\
		 \frac{1}{57} & \langle
		\left( 4O^{(4)}_{3,n-1}-52 O^{(4)}_{1,n-1} +57 O^{(4)}_{8,n-1} (n-1)-3 O^{(4)}_{4,n-1}+ 3 O^{(4)}_{7,n-1} (n-1)\right.
		\nonumber\\ & \qquad\left. -5 O^{(4)}_{6,n-1} (n-1)^2 \right)
		 \mathcal{O}^\dagger_2 \mathcal{O}^\dagger_n \rangle /(n-1)
		\nonumber\\ & = \frac{235 n^{-\frac{n+3}{n-1}}}{228 (n-1)} \sqrt{\frac{228(n-1)^2}{235}}C^{\star}_{n,2,n-1}
	\label{eq:3pfs_k=4_down_firstlast}
	\end{align}
where $C^{\star}_{n,2,n-1}$ is the leading coefficient in (\ref{eq:twist_down_4pf}). Adding up the squares of (\ref{eq:3pfs_k=4_down_firstlast}) gives
	\begin{equation}
		\frac{1}{3} n^{-\frac{6 n+2}{n-1}} \left(15 n^6+94 n^5+256 n^4+348 n^3+256 n^2+94 n+15\right) \left(C^{\star}_{n,2,n-1}\right)^2
	\end{equation}
which matches the order $w^{4/(n-1)}$ coefficient in (\ref{eq:twist_down_4pf}).

	Operators contributing at level $k=5$ and their norms are
	\begin{align}
		O^{(5)}_{1,n} &= \left( {J'}^3_{-\frac{1}{n}} \right)^5 O_n
		& \langle |O^{(5)}_{1,n}|^2 \rangle &=   \frac{15}{4}
		& O^{(5)}_{9,n} &=  {L'}_{-\frac{5}{n}}   O_n
		& \langle |O^{(5)}_{9,n}|^2 \rangle &=  \frac{60}{n^2}	
		\nonumber\\
		O^{(5)}_{2,n} &=  {J'}^3_{-\frac{2}{n}} \left( {J'}^3_{-\frac{1}{n}} \right)^3 O_n
		& \langle |O^{(5)}_{2,n}|^2 \rangle &=  \frac{3}{4}
		& O^{(5)}_{10,n} &=  {J'}^3_{-\frac{3}{n}} {L'}_{-\frac{2}{n}}    O_n
		& \langle |O^{(5)}_{10,n}|^2 \rangle &=  \frac{8}{2n^2}	
		\nonumber\\
		O^{(5)}_{3,n} &=  {J'}^3_{-\frac{3}{n}} \left( {J'}^3_{-\frac{1}{n}} \right)^2   O_n
		&\langle |O^{(5)}_{3,n}|^2 \rangle &=   \frac{3}{4}
		& O^{(5)}_{11,n} &= \left( {J'}^3_{-\frac{1}{n}} \right)^3 {L'}_{-\frac{2}{n}}    O_n
		& \langle |O^{(5)}_{11,n}|^2 \rangle &=  \frac{45}{8n^2}	
		\nonumber\\
		O^{(5)}_{4,n} &=  {J'}^3_{-\frac{4}{n}}  {J'}^3_{-\frac{1}{n}}    O_n
		& \langle |O^{(5)}_{4,n}|^2 \rangle &= 1
		& O^{(5)}_{12,n} &=  {J'}^3_{-\frac{2}{n}} {J'}^3_{-\frac{1}{n}} {L'}_{-\frac{2}{n}}    O_n
		& \langle |O^{(5)}_{12,n}|^2 \rangle &=  \frac{2}{n^2}	
		\nonumber\\
		O^{(5)}_{5,n} &=  \left( {J'}^3_{-\frac{2}{n}} \right)^2 {J'}^3_{-\frac{1}{n}}    O_n
		& \langle |O^{(5)}_{5,n}|^2 \rangle &=  1
		& O^{(5)}_{13,n} &= {J'}^3_{-\frac{1}{n}}  \left({L'}_{-\frac{2}{n}} \right)^2   O_n
		& \langle |O^{(5)}_{13,n}|^2 \rangle &=  \frac{61}{2n^4}	
		\nonumber\\
		O^{(5)}_{6,n} &=  {J'}^3_{-\frac{3}{n}}  {J'}^3_{-\frac{2}{n}}    O_n
		& \langle |O^{(5)}_{6,n}|^2 \rangle &=  \frac{3}{2}
		& O^{(5)}_{14,n} &= \left( {J'}^3_{-\frac{1}{n}}\right)^2  {L'}_{-\frac{3}{n}}    O_n
		& \langle |O^{(5)}_{14,n}|^2 \rangle &=  \frac{8}{n^2}	
		\nonumber\\
		O^{(5)}_{7,n} &=  {J'}^3_{-\frac{5}{n}}    O_n
		& \langle |O^{(5)}_{7,n}|^2 \rangle &=  \frac{30}{12}
		& O^{(5)}_{15,n} &= {J'}^3_{-\frac{2}{n}} {L'}_{-\frac{3}{n}}    O_n
		& \langle |O^{(5)}_{15,n}|^2 \rangle &=  \frac{14}{n^2}	
		\nonumber\\
		O^{(5)}_{8,n} &=  L'_{-\frac{3}{n}} L'_{-\frac{2}{n}}    O_n
		& \langle |O^{(5)}_{8,n}|^2 \rangle &=  \frac{72}{n^4}
		& O^{(5)}_{16,n} &=  {J'}^3_{-\frac{1}{n}}{L'}_{-\frac{4}{n}}    O_n
		& \langle |O^{(5)}_{16,n}|^2 \rangle &=  \frac{33}{2n^2}\,.	
	\end{align}
	\vskip-0.15truein
The twist up 3-point functions at level $k=5$ are
	\begin{align}
		 \langle O^{(5)}_{9,n+1}       \mathcal{O}^\dagger_2 \mathcal{O}^\dagger_n \rangle
		&= \frac{(-n)^{\frac{n+6}{n+1}} \left(12 n^3-47 n^2+47 n-12\right)}{n^5 (n+1)} \sqrt{\frac{(n+1)^2}{60}}C_{n,2,n+1}
		\nonumber\\
		 \langle O^{(5)}_{7,n+1}       \mathcal{O}^\dagger_2 \mathcal{O}^\dagger_n \rangle
		&= \frac{(-n)^{\frac{5}{n+1}} \left(-12 n^3+47 n^2-47 n+12\right)}{6 n^4} \sqrt{\frac{2}{5}}C_{n,2,n+1}
		\nonumber\\
		 \langle O^{(5)}_{10,n+1}       \mathcal{O}^\dagger_2 \mathcal{O}^\dagger_n \rangle
		&= -\frac{3 (n-1) (-n)^{\frac{2-3 n}{n+1}}}{2 (n+1)} \sqrt{\frac{2(n+1)^2}{9}}C_{n,2,n+1}
		\nonumber\\
		 \langle O^{(5)}_{15,n+1}       \mathcal{O}^\dagger_2 \mathcal{O}^\dagger_n \rangle
		&= -\frac{7 (n-1) (-n)^{\frac{2-3 n}{n+1}}}{3 (n+1)} \sqrt{\frac{(n+1)^2}{14}}C_{n,2,n+1}
		\nonumber
	\end{align}
	\vskip-0.3truein
	\begin{align}
		& \langle \left( O^{(5)}_{8,n+1} - \frac{2}{5 (n+1)}O^{(5)}_{9,n+1} \right)       \mathcal{O}^\dagger_2 \mathcal{O}^\dagger_n \rangle
		= \frac{10 (n-1) (-n)^{\frac{2 n+7}{n+1}}}{n^5 (n+1)^2} \sqrt{\frac{5(n+1)^4}{312}}C_{n,2,n+1}
		\nonumber\\
		& \langle \left(O^{(5)}_{6,n+1}+\frac{n+1}{52} (O^{(5)}_{8,n+1} (n+1)-3 O^{(5)}_{9,n+1}) \right)        \mathcal{O}^\dagger_2 \mathcal{O}^\dagger_n \rangle
		= \frac{23 (n-1) (-n)^{\frac{2 n+7}{n+1}}}{52 n^5} \sqrt{\frac{52}{69}}C_{n,2,n+1} \,.
		\label{eq:3pfs_k=5_up_firstlast}
	\end{align}
	\vskip-0.15truein
\noindent Adding up the squares of (\ref{eq:3pfs_k=5_up_firstlast}) gives
	\begin{equation}
		\frac{2}{3} (n-1)^2 (-n)^{\frac{2-8n}{n+1}}(6 n^4-35 n^3+67 n^2-35 n+6)
	\end{equation}
which matches the order $w^{5/(n+1)}$ coefficient of (\ref{CPCPTwistUpExpand}).

Finally, the twist down 3-point functions at level $k=5$ are
	\begin{align}
		\langle O^{(5)}_{1,n-1} \mathcal{O}^\dagger_2 \mathcal{O}^\dagger_n \rangle
		&= \frac{1}{16} n^{\frac{2 n+3}{1-n}} \left(16 n^2+40 n+15\right) \sqrt{\frac{4}{15}}C^{\star}_{n,2,n-1}
\nonumber\\
		\langle O^{(5)}_{7,n-1} \mathcal{O}^\dagger_2 \mathcal{O}^\dagger_n \rangle
		&= \frac{1}{24} n^{-\frac{5}{n-1}-4} \left(24 n^4+202 n^3+457 n^2+342 n+72\right) \sqrt{\frac{2}{5}}C^{\star}_{n,2,n-1}
\nonumber\\
		\langle O^{(5)}_{9,n-1} \mathcal{O}^\dagger_2 \mathcal{O}^\dagger_n \rangle
		&= \frac{n^{-\frac{5}{n-1}-4} \left(24 n^4+202 n^3+457 n^2+342 n+72\right)}{6 (n-1)} \sqrt{\frac{(n-1)^2}{60}}C^{\star}_{n,2,n-1}
\nonumber\\
		\langle O^{(5)}_{10,n-1} \mathcal{O}^\dagger_2 \mathcal{O}^\dagger_n \rangle
		&= \frac{n^{-\frac{5}{n-1}-3} \left(8 n^3+46 n^2+73 n+30\right)}{8 (n-1)} \sqrt{\frac{2(n-1)^2}{9}}C^{\star}_{n,2,n-1}
\nonumber\\
		\langle O^{(5)}_{15,n-1} \mathcal{O}^\dagger_2 \mathcal{O}^\dagger_n \rangle
		&= \frac{n^{-\frac{5}{n-1}-3} \left(12 n^3+70 n^2+109 n+48\right)}{6 (n-1)} \sqrt{\frac{(n-1)^2}{14}}C^{\star}_{n,2,n-1}\nonumber
	\end{align}
	\vskip-0.3truein
		\begin{align}
		\langle \left( O^{(5)}_{6,n-1} - \frac{n-1}{20	}O^{(5)}_{9,n-1} \right) \mathcal{O}^\dagger_2 \mathcal{O}^\dagger_n \rangle
		&= \frac{n^{\frac{3 n+2}{1-n}}}{120}  \left(96 n^3+470 n^2+665 n+270\right) \sqrt{\frac{20}{27}}C^{\star}_{n,2,n-1}
\nonumber\\
		\langle \left( O^{(5)}_{3,n-1}-\frac{ n-1}{6}O^{(5)}_{10,n-1}  \right) \mathcal{O}^\dagger_2 \mathcal{O}^\dagger_n \rangle
		&= \frac{5}{48} n^{-\frac{5}{n-1}-3} \left(8 n^3+34 n^2+31 n+6\right) \sqrt{\frac{8}{5}}C^{\star}_{n,2,n-1}
\nonumber\\
		\langle \left( O^{(5)}_{5,n-1} - \frac{ n-1}{7}O^{(5)}_{15,n-1}  \right) \mathcal{O}^\dagger_2 \mathcal{O}^\dagger_n \rangle
		&= \frac{5}{28} n^{-\frac{5}{n-1}-3} \left(4 n^3+14 n^2+13 n+2\right) \sqrt{\frac{7}{5}}C^{\star}_{n,2,n-1}
\nonumber\\
		\langle \left( O^{(5)}_{11,n-1}-\frac{1}{n-1}O^{(5)}_{1,n-1}  \right) \mathcal{O}^\dagger_2 \mathcal{O}^\dagger_n \rangle
		&= \frac{5 n^{\frac{2 n+3}{1-n}} (4 n+3)}{16 (n-1)} \sqrt{\frac{8(n-1)^2}{15}}C^{\star}_{n,2,n-1} \label{eq:3pfs_k=5_down_first}
	\end{align}
	
\noindent and

\noindent
\begin{align}
		&\langle \left(  O^{(5)}_{8,n-1} + \frac{8  O^{(5)}_{6,n-1}}{9 (n-1)^2} - \frac{4  O^{(5)}_{9,n-1}}{9 (n-1)}  \right) \mathcal{O}^\dagger_2 \mathcal{O}^\dagger_n \rangle
		\nonumber\\  & = \frac{n^{\frac{3 n+2}{1-n}} \left(84 n^3+508 n^2+817 n+351\right)}{27 (n-1)^2} \sqrt{\frac{3(n-1)^4}{184}}C^{\star}_{n,2,n-1}
\nonumber\\
		\frac{1}{368} & \langle \left(  368 O^{(5)}_{2,n-1}-8 O^{(5)}_{6,n-1} +(n-1) (4 O^{(5)}_{9,n-1} -9 O^{(5)}_{8,n-1} (n-1))  \right) \mathcal{O}^\dagger_2 \mathcal{O}^\dagger_n \rangle
		\nonumber\\ & =  \frac{5 n^{\frac{3 n+2}{1-n}} \left(204 n^3+616 n^2+361 n-15\right)}{1104} \sqrt{\frac{736}{525}}C^{\star}_{n,2,n-1}
\nonumber\\
		\frac{1}{105} & \langle \left( 4 O^{(5)}_{2,n-1}+105 O^{(5)}_{4,n-1}+6 O^{(5)}_{6,n-1}-2 O^{(5)}_{8,n-1} (n-1)^2-3 O^{(5)}_{9,n-1} (n-1)  \right) \mathcal{O}^\dagger_2 \mathcal{O}^\dagger_n \rangle
		\nonumber\\ &= \frac{1}{42} n^{\frac{3 n+2}{1-n}} \left(38 n^3+217 n^2+292 n+95\right) \sqrt{\frac{21}{19}}C^{\star}_{n,2,n-1} \nonumber\\
		\frac{1}{(n-1)^2} & \langle \left( O^{(5)}_{1,n-1}-2O^{(5)}_{11,n-1} (n-1)+O^{(5)}_{13,n-1} (n-1)^2-2 O^{(5)}_{3,n-1}  \right) \mathcal{O}^\dagger_2 \mathcal{O}^\dagger_n \rangle
		\nonumber\\ &= \frac{5 n^{\frac{3 n+2}{1-n}} \left(12 n^2+37 n+12\right)}{16 (n-1)^2} \sqrt{\frac{4(n-1)^4}{65}}C^{\star}_{n,2,n-1}
\nonumber\\
		\frac{1}{114(n-1)}& \langle \left( 2 (n-1) (57 O^{(5)}_{12,n-1}+O^{(5)}_{9,n-1})-104 O^{(5)}_{2,n-1}+6 O^{(5)}_{4,n-1}-4 O^{(5)}_{6,n-1} \right.
		\nonumber\\ & \qquad \left.-5 O^{(5)}_{8,n-1} (n-1)^2  \right) \mathcal{O}^\dagger_2 \mathcal{O}^\dagger_n \rangle
		\nonumber\\ &= \frac{5 n^{\frac{2 n+3}{1-n}} (68 n+107)}{342 (n-1)} \sqrt{\frac{57(n-1)^2}{65}}C^{\star}_{n,2,n-1}
\nonumber\\
		\frac{1}{52(n-1)} & \langle \left( 2 (n-1) (5 O^{(5)}_{12,n-1}+26 O^{(5)}_{14,n-1}+O^{(5)}_{9,n-1})-104 O^{(5)}_{2,n-1}+6 O^{(5)}_{4,n-1} \right.
		\nonumber\\ & \qquad \left. -4 O^{(5)}_{6,n-1}-5 O^{(5)}_{8,n-1}(n-1)^2  \right) \mathcal{O}^\dagger_2 \mathcal{O}^\dagger_n \rangle
		\nonumber\\ &= \frac{235 n^{\frac{2 n+3}{1-n}} (n+1)}{78 (n-1)} \sqrt{\frac{52(n-1)^2}{235}}C^{\star}_{n,2,n-1}
\nonumber\\
		-\frac{1}{13(n-1)} & \langle \left( 6 O^{(5)}_{1,n-1}-12 O^{(5)}_{11,n-1} (n-1)+6 O^{(5)}_{13,n-1} (n-1)^2-13 O^{(5)}_{16,n-1} (n-1) \right.
		\nonumber\\ & \qquad \left. +14 O^{(5)}_{3,n-1}+13 O^{(5)}_{5,n-1} \right) \mathcal{O}^\dagger_2 \mathcal{O}^\dagger_n \rangle
		\nonumber\\ & = \frac{235 n^{\frac{3 n+2}{1-n}} (n+1)^2}{52 (n-1)} \sqrt{\frac{26(n-1)^2}{235}}C^{\star}_{n,2,n-1} \,.\label{eq:3pfs_k=5_down_last}
	\end{align}
Adding the squares of (\ref{eq:3pfs_k=5_down_first}) and (\ref{eq:3pfs_k=5_down_last}) gives the result
	\begin{align}
	&\frac{1}{3} n^{\frac{8 n+2}{1-n}} \left(18 n^8+171 n^7+706 n^6+1561 n^5+2022 n^4+1561 n^3+706 n^2+171 n+18\right)
	\nonumber\\ & \quad \times \left(C^{\star}_{n,2,n-1}\right)^2
	\end{align}
which matches the order $w^{5/(n-1)}$ coefficient of (\ref{eq:twist_down_4pf}).


\end{document}